\documentclass[11pt,a4paper]{article}
\pdfoutput=1
\usepackage{jcappub}
\usepackage{aasmacros}

\usepackage{graphicx}
\usepackage{amsmath}
\usepackage{amssymb}
\usepackage{rotating} 
\usepackage{xspace}
\usepackage{color}
\usepackage{booktabs}
\usepackage[normalem]{ulem}

\newcommand{\Msun}{{\rm M}_\odot}

\title{The dark matter component of the Gaia radially anisotropic substructure}

\author[a, b]{Nassim Bozorgnia,}
\author[c]{Azadeh Fattahi,}
\author[c]{Carlos S. Frenk,}
\author[b, d]{Andrew Cheek,}
\author[b]{David G.~Cerde\~no,}
\author[e, f]{Facundo A. G\'{o}mez,}
\author[g]{Robert J. J. Grand,}
\author[h]{and Federico Marinacci}

\affiliation[a]{York University, Department of Physics and Astronomy,\\
4700 Keele Street, Toronto, Ontario M3J 1P3, Canada}
\affiliation[b]{Institute for Particle Physics Phenomenology, Department of Physics,\\
Durham University, Durham DH1 3LE, UK} 
\affiliation[c]{Institute for Computational Cosmology, Durham University,\\
South Road, Durham DH1 3LE, UK}
\affiliation[d]{Centre for Cosmology, Particle Physics and Phenomenology (CP3),\\
Universit\'e catholique de Louvain, Chemin du Cyclotron 2,\\
B-1348 Louvain-la-Neuve, Belgium }
\affiliation[e]{Instituto de Investigaci\'on Multidisciplinar en Ciencia y Tecnolog\'ia, \\
Universidad de La Serena, Ra\'ul Bitr\'an 1305, La Serena, Chile}
\affiliation[f]{Departamento de F\'isica y Astronom\'ia, Universidad de La Serena,\\
Av. Juan Cisternas 1200 Norte, La Serena, Chile}
\affiliation[g]{Max-Planck-Institut f\"{u}r Astrophysik,\\
Karl-Schwarzschild-Str. 1, D-85748, Garching, Germany}
\affiliation[h]{Department of Physics and Astronomy, University of Bologna,\\
via Gobetti 93/2, 40129 Bologna, Italy}

\emailAdd{nassimb@yorku.ca}

\abstract{We study the properties of the dark matter component of the radially anisotropic stellar population recently identified in the Gaia data, using magneto-hydrodynamical simulations of Milky Way-like halos from the Auriga project. We identify 10  simulated galaxies that approximately match the rotation curve and stellar mass of the  Milky Way. Four of these  have an anisotropic stellar population reminiscent of the Gaia structure. We find an anti-correlation between the  dark matter mass fraction of this population in the Solar neighbourhood and its orbital anisotropy. We estimate  the local dark matter density and velocity distribution for halos with and without the anisotropic stellar population, and use them to simulate the signals expected in future xenon and germanium direct detection experiments. We find that a generalized Maxwellian distribution fits the dark matter halo integrals of the Milky Way-like halos containing the radially anisotropic stellar population. For dark matter particle masses below approximately 10 GeV, direct detection exclusion limits for the simulated halos with the anisotropic stellar population show a mild shift towards smaller masses compared to the commonly adopted Standard Halo Model.}

\arxivnumber{IPPP/19/75, CP3-19-45}

\begin{document}
\maketitle

\section{Introduction}
\label{sec:introduction}

The second data release from the \emph{Gaia} satellite~\cite{Brown:2018dum} has revolutionised our understanding of the Milky Way (MW) halo. Among many interesting findings, Gaia has set constraints on the mass and shape of the dark matter (DM) halo~\cite{2019A&A...621A..56P, Wegg:2018voc, 2019ApJ...873..118W}, the MW halo potential~\cite{2019MNRAS.486.2995M}, and the Galactic escape speed~\cite{2018A&A...616L...9M, 2019MNRAS.485.3514D, 2019MNRAS.487L..72G}. One of the recent important discoveries made using Gaia data was a prominent population of stars in the inner stellar halo (within $\sim 10$~kpc from the Sun) which have a high radial velocity anisotropy~\cite{Belokurov2019, Helmi2018,Myeong2018} (see also \cite{Carollo:2007xh,Mackereth2019,Navarro2011}). This population has been named the \emph{Gaia sausage} in ref.~\cite{Belokurov2019} and \emph{Gaia Enceladus} in ref.~\cite{Helmi2018}. However it is not clear whether the Gaia sausage and Enceladus are the same structure~\cite{2019MNRAS.488.1235M}. The two structures are clearly linked and overlap significantly: Gaia Enceladus appears to include parts of the Gaia sausage as well as debris from a dwarf galaxy less massive than the sausage progenitor, called \emph{Sequoia}~\cite{2019MNRAS.488.1235M, 2020arXiv200205740E}.

The structure we study in this work is defined to be similar to that of ref.~\cite{Belokurov2019}. This population dominates the inner stellar halo and consists of relatively metal rich stars ($[{\rm Fe}/{\rm H}]\sim-1$). It has been shown, using both idealized and cosmological simulations, that this population originated from a relatively massive dwarf galaxy ($M_\star \sim 10^9~\Msun$) on a radial orbit, which merged with the MW around $10$~Gyr ago~\cite{Belokurov2019,Mackereth2019,Fattahi2019} (see also \cite{Brook2003,Meza2005}). Using the Auriga cosmological simulations, ref.~\cite{Fattahi2019} showed that even though such a merger contributes significantly to the build-up of the inner stellar halo of the MW, DM brought in by the merging dwarf galaxy typically makes up less than $\sim10\%$ of the final mass in the inner 20 kpc.    
In this work, we use the acronym \emph{GRASP} (Gaia Radially Anisotropic Stellar Population) for referring to this population of stars.

A pressing question regarding the GRASP, is the properties of its unknown DM component
(i.e.~DM originating from the GRASP progenitor), and its implications for the interpretation of DM direct detection results. In particular, DM direct detection event rates are sensitive to the DM velocity distribution in the Solar neighbourhood, and variations in this distribution can lead to large uncertainties in the interpretation of direct detection data. Given the high radial velocity anisotropy of the stars in the GRASP and their high mass fraction in the inner halo, it is especially important to study how its DM component may modify the commonly adopted Standard Halo Model (SHM)~\cite{Drukier:1986tm} assumption in the analysis of direct detection data~\cite{Necib:2018iwb, Evans:2018bqy, Buch:2019aiw}.  

In the SHM, the DM halo is assumed to be isothermal and the DM velocity distribution is an isotropic Maxwellian distribution with a peak speed equal to the local circular speed, usually set to $v_c=220$~km~s$^{-1}$. The true DM distribution could however be different from the SHM, modifying the exclusion limits set by direct detection experiments in the plane of DM mass and interaction cross section~\cite{Green:2000jg,Green:2002ht,Vogelsberger:2008qb,Kuhlen:2009vh, Lisanti:2010qx, Fairbairn:2012zs, Bozorgnia:2013pua}. Uncertainties in the parameters of the SHM could also significantly impact DM direct detection limits~\cite{McCabe:2010zh, Green:2010gw, Green:2011bv, Benito:2016kyp, Green:2017odb, Wu:2019nhd}.

High resolution cosmological simulations of galaxy formation, which include baryonic physics, have reached significant agreement with observations. These more realistic simulations are a powerful tool to provide insight regarding the local DM distribution. In particular, recent studies have found that the local DM velocity distribution of MW analogues in the EAGLE/APOSTLE, MaGICC, and the Sloane {\it et al.} hydrodynamic simulations agrees well with a Maxwellian velocity distribution\footnote{The best fit peak speed of the  Maxwellian distribution and the local circular speed of the simulated halo may however be different, as seen in figure 1 of ref.~\cite{Bozorgnia:2017brl}.}~\cite{Bozorgnia:2016ogo, Kelso:2016qqj, Sloane:2016kyi, Bozorgnia:2017brl}. However, these works were carried out before the discovery of the GRASP, and it is timely to study and understand how the DM component of the GRASP can modify the SHM.

Very recently, a number of studies were carried out to model the local velocity distribution of the DM component of the GRASP~\cite{Necib:2018iwb, Evans:2018bqy}, using a combination of observational data, simulations, and theoretical modelling. Ref.~\cite{Necib:2018iwb} used the stellar distributions from SDSS-Gaia DR2 to model the local DM velocity distribution. This was motivated by the results from the ERIS and FIRE-2 simulations which suggest that old metal-poor stars trace the DM accreted from the oldest luminous mergers~\cite{Herzog-Arbeitman:2017fte, Necib:2018igl}, and intermediate metallicity stars trace the DM in debris flow accreted from younger mergers~\cite{Necib:2018igl}. Ref.~\cite{Necib:2018iwb} characterized the local DM halo as a two component distribution, with the total local DM speed distribution having fewer particles in the high speed tail compared to the SHM. However, a non-negligible part of the local DM halo originates from smooth accretion as well as dark substructures~\cite{Wang2011, Sawala2015, Angulo:2009hf}, which were not taken into account in the results of refs.~\cite{Herzog-Arbeitman:2017fte, Necib:2018igl, Necib:2018iwb}. In fact, the total local DM distribution, regardless of its origin, is required in the analysis of direct detection data. Our recent analysis of the Auriga simulations shows no correlations between the total local DM velocity distribution and the velocity distribution of old or metal-poor stars~\cite{Bozorgnia:2018pfa}.

Ref.~\cite{Evans:2018bqy} provides an analytic velocity distribution for the DM in our Solar neighbourhood including the DM component originating from the GRASP as well as the DM belonging to the isotropic halo. The authors argue that the fraction of the local DM in the GRASP is between 10\% and 30\%, and model the DM velocity distribution as a linear combination of an isotropic Maxwellian distribution for the smooth halo and an anisotropic Gaussian distribution for the GRASP DM component. The radial velocity anisotropy of the DM particles belonging to the GRASP is set equal to the anisotropy of stars in the GRASP, i.e.~$\beta=0.9$ (see eq.~\eqref{eq:beta} for the definition of the anisotropy parameter, $\beta$). The DM density profile of the GRASP is modelled as $\rho(r) \propto r^{-3}$ in the Solar neighbourhood, and using a limit on the ellipticity of the Galactic disc leads to an upper limit on the fraction of DM due to the GRASP of 20\% in the Solar neighbourhood. Hence, ref.~\cite{Evans:2018bqy} introduces the SHM$^{++}$ which includes updated Galactic parameters, and the combination of an isotropic and anisotropic velocity distributions for the DM in the Solar neighbourhood. The local DM velocity distribution in the SHM$^{++}$  is shifted to higher speeds and has a higher peak height, compared to the SHM.

In this work, we use the Auriga magneto-hydrodynamical simulations of galaxy formation~\cite{Grand:2016mgo} to identify the DM component of the GRASP, extract the local DM distribution from the simulated halos, and study its implications for DM direct detection. In section~\ref{sec:sims} we provide the details of the simulations we use, and in section~\ref{sec:MilkyWay} we review the criteria we use to identify simulated MW-like galaxies. In section~\ref{sec:properties} we discuss the properties of the DM debris originating from the GRASP progenitor. In sections~\ref{sec:density} and \ref{sec:velocities} we present the DM density profiles and velocity distributions for halos with and without the GRASP. The implications of the DM component of the GRASP for direct detection are discussed in section \ref{sec:dirdet}, and we conclude in section~\ref{sec:conclusions}. 

%*******************************%
\section{Auriga Simulations}
\label{sec:sims}

We use magneto-hydrodynamical simulations of MW-size halos from the Auriga project~\citep{Grand:2016mgo}. This simulation suite includes cosmological zoom-in simulations of 30, relatively isolated halos with virial mass\footnote{Virial quantities are defined here as those corresponding to a spherical radius where the mean enclosed matter density is 200 times the critical density of the Universe. We denote these parameters with a 200 subscript.} $M_{200}\sim10^{12}\,\Msun$, selected from a $100^3$~Mpc$^3$ periodic box (L100N1504) from the EAGLE project~\cite{Schaye2015,Crain2015}. The simulations were performed using the moving-mesh code Arepo~\citep{Springel2010} and are complemented by a galaxy formation subgrid model which includes metal cooling, star formation, supernovae feedback, and background UV/X-ray photoionisation radiation (see ref.~\cite{Grand:2016mgo} for full details). 

In this work we use the fiducial resolution level (Level 4) of the simulations with DM particle mass of approximately $3\times 10^5~\Msun$, baryonic mass resolution of $5\times10^4~\Msun$, and Plummer equivalent gravitational softening of $370$~pc~\citep{Power2003,Jenkins2013}. We discard halos Au11 and Au20, since they are under-going a merger. DM halos and bound structures in the simulations are found using the friend-of-friends algorithm and SUBFIND~\citep{Davis1985,Springel2001a}, respectively. The cosmological parameters used by the simulations are, according to Planck-2015~\citep{Planck2015}: $\Omega_{m}=0.307$, $\Omega_{\rm bar}=0.048$, $H_0=67.77~{\rm km~s^{-1}~Mpc^{-1}}$.

All the simulated halos have their dark-matter-only (DMO) counterparts which share the same initial conditions as the  magneto-hydrodynamical runs, but galaxy formation processes (and magneto-hydrodynamics) are ignored and all the particles are collisionless and interact only gravitationally.

The position and velocity reference frame (i.e.~the centre) of the simulated halos are calculated using an iterative, shrinking sphere method on DM particles\footnote{We have checked that including stars gives a similar result.}. In this method, we start with the virial radius of the halo and shrink the radius by 2.5\% at each step, until 1000 particles remain, which roughly corresponds to a sphere with radius similar to the gravitational softening. The stellar masses of the MW-size halos are calculated from the stars within a spherical radius of $30$~kpc from the centre. We use the angular momentum of stars within the inner 10 kpc to define the $z$-axis, hence \emph{disk}, of the simulated halos. We adopt the  same coordinate transformation for the DMO halos as their magneto-hydrodynamical counterparts.

The highly radial components of the stellar halos in the Auriga galaxies, i.e.~GRASP-like components, were defined in detail in ref.~\cite{Fattahi2019}. In summary, the velocity space of metal rich ($[{\rm Fe}/{\rm H}]\sim -1$) \emph{accreted} stars above and below the disk ($|z|\sim10$~kpc) was decomposed into $2-3$ Gaussian components, adopting the same technique used in observations~\citep{Belokurov2019}. The components which are highly anisotropic ($\beta>0.8$) and dominate the accreted stellar halo in the inner regions (mass contribution $>50\%$), were defined as \emph{GRASP-like}. It was found that 10 out of 28 Auriga halos have such a component. 

Stars associated to the GRASP component of a given Auriga halo can have multiple progenitors. The main progenitor is identified as the accreted galaxy which contributed the most mass to it. In this work we consider star and DM particles which were bound to the GRASP main progenitors at infall; i.e.~when the dwarf galaxy crosses the virial radius of the MW progenitor for the first time.         

%*******************************%
\subsection{Identifying Milky Way analogues}
\label{sec:MilkyWay}

To make accurate predictions for the local DM distribution from simulations, we need to identify  simulated  halos which closely resemble the MW. The virial mass of the Auriga halos is in the range of $M_{\rm 200}=[0.93 - 1.91] \times 10^{12}~\Msun$~\cite{Grand:2016mgo}, agreeing well with the recent estimates for the MW halo mass range from observations (see ref.~\cite{Callingham2019} and references therein for a compilation of recent measurements).  
Following ref.~\cite{Bozorgnia:2016ogo}, we identify the MW analogues in Auriga by requiring that: (i) the total stellar mass of the simulated galaxy falls within the 3$\sigma$ range of the  MW stellar mass  derived from observations, $4.5 \times10^{10}<M_{*}/\Msun<8.3 \times10^{10}$~\cite{McMillan:2011wd}, and (ii) the rotation curves of the simulated halos fit well the observed MW rotation curve obtained from ref.~\cite{Iocco:2015xga}. The details of our selection procedure are given in refs.~\cite{Bozorgnia:2016ogo} and \cite{Calore:2015oya}.

\begin{figure}[t!]
\begin{center}
  \includegraphics[width=0.7\textwidth]{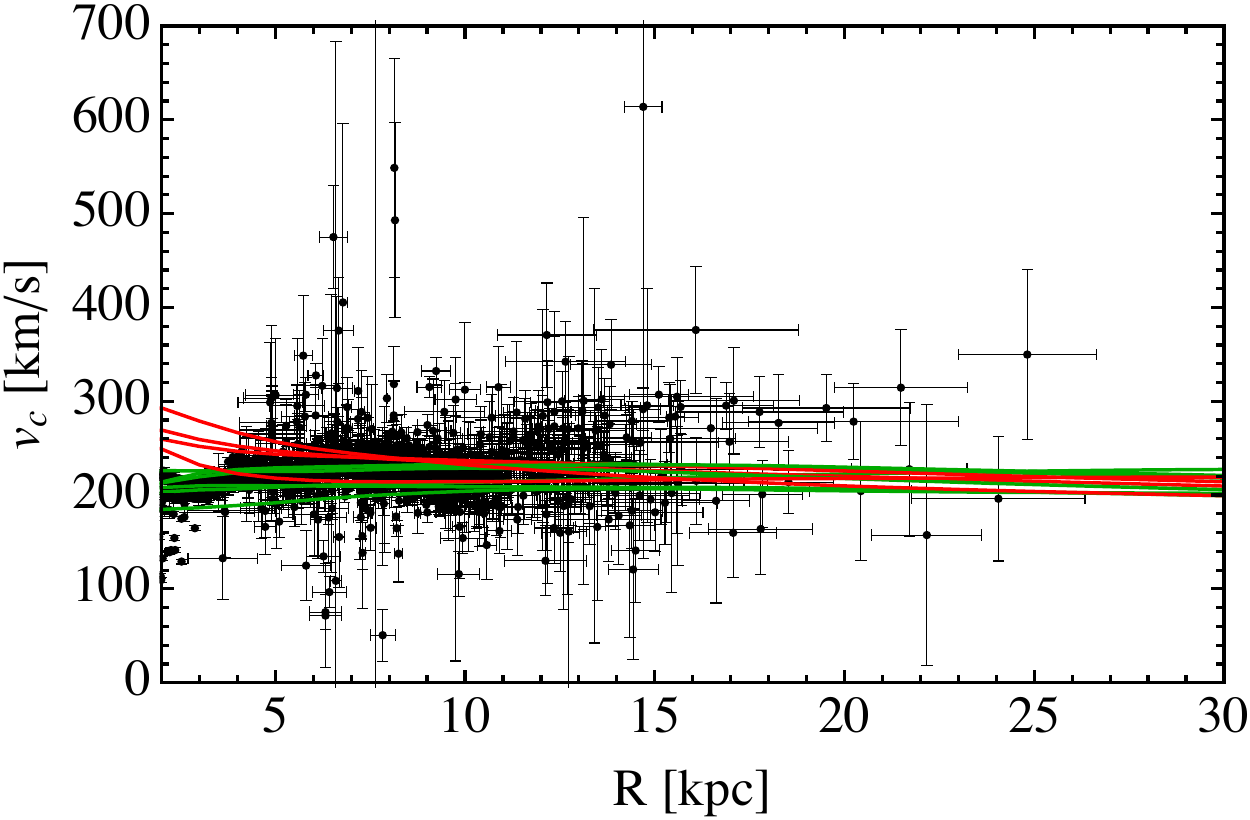}
\caption{Observed MW rotation curves from ref.~\cite{Iocco:2015xga} (black data points and their $1\sigma$ error bars) and rotation curves of Auriga MW analogues with (red curves) and without (green curves) the GRASP.}
\label{fig:vc}
\end{center}
\end{figure}

From the 28 Auriga halos which are not currently undergoing a merger, 16 halos have the correct MW stellar mass from criterion (i). From those 16 halos, we select a set of 10 halos which provide the best agreement with the observed MW rotation curve\footnote{To compute the goodness of fit, we consider rotation curve measurements for Galactocentric distances $R > 2.5$~kpc, since the gravitational potential of the Galactic bulge becomes important at smaller radii and the tracers in ref.~\cite{Iocco:2015xga} would not have circular orbits. The details of the fitting procedure are given in ref.~\cite{Calore:2015oya}.}. Figure~\ref{fig:vc} shows the circular velocity, $v_c$, as a function of the Galactocentric distance, $R$, for the kinematical data from ref.~\cite{Iocco:2015xga} (black data points and $1\sigma$ error bars) and the 10  Auriga MW-like simulated halos. The rotation curves for halos with and without the GRASP are shown as red and green curves, respectively. As it can be seen from figure~\ref{fig:vc}, the rotation curves of our final set of Auriga MW-like halos provide a reasonable agreement with the observed MW rotation curve. Notice that the rotation curves of some halos show an increase towards the centre of the galaxy. This feature which occurs in a number of Auriga halos is independent of the halos having the GRASP, and is due to their centrally concentrated stellar distributions~\cite{Grand:2016mgo}. 

The virial mass and the total stellar mass for the 10 Auriga halos which satisfy our criteria are listed in table \ref{tab:MWlike}. Out of these 10 halos, the four halos Au5, Au9, Au22, and Au24 have the GRASP component (i.e.~they meet the criteria described in ref.~\cite{Fattahi2019}, as explained above), and are labeled with a $(\star)$ symbol. 

  \begin{table}[h!]
    \centering
    \begin{tabular}{|c|c|c|}
      \hline
      Halo Name  & $M_{\rm 200}~[\times 10^{12} \, \Msun]$&  $M_\star~[\times 10^{10} \, \Msun]$ \\
      \hline
      Au2 & 1.91 & 7.65 \\
      Au4 & 1.41 & 7.54 \\
      ~~~~~Au5 ($\star$) & 1.19 & 6.88  \\
      Au7 &  1.12 & 5.27 \\
      ~~~~~Au9 ($\star$) & 1.05 & 6.20 \\
      Au12 & 1.09 & 6.29 \\
      Au19 & 1.21 &  5.72 \\
      Au21 & 1.45 & 8.02\\
      ~~~~~Au22 ($\star$)  & 0.93 & 6.10 \\
      ~~~~~Au24 ($\star$) & 1.49 & 7.07\\
      \hline
    \end{tabular}
\caption{The virial mass and stellar mass of the Auriga MW-like halos. The halos with the GRASP are identified with a ($\star$) symbol.}
    \label{tab:MWlike}
  \end{table}

%*******************************%

\section{Dark matter debris from the GRASP progenitor}
\label{sec:properties}

In this section, we explore the properties of the DM debris originating from the GRASP progenitor in the Solar neighbourhood. As discussed in the previous section, 10 Auriga halos have the GRASP component, and only four of them are MW analogues according to our criteria defined in section~\ref{sec:MilkyWay}. However, here we discuss the properties of all 10 halos with the GRASP. 

Since direct detection experiments are sensitive to the \emph{local} DM distribution, for almost all our analyses in this paper (other than the computation of the DM density profiles in section~\ref{sec:density}) we consider the DM particles in a torus with a square cross section (i.e.~a cylindrical shell) aligned with the stellar disc, with radius $7 \leq \rho \leq 9$~kpc from the Galactic centre, and height $|z| \leq 2$~kpc, with respect to the Galactic plane. We consider a DM particle in the torus to belong to the GRASP, if it was bound to the GRASP progenitor at infall. The total number of DM particles belonging to the GRASP in the torus region is in the range of $[69 - 5776]$ depending on the halo. For the four MW-like halos with the GRASP, the range is $[69 - 2398]$. In the second column of table~\ref{tab:sausageprop}, we present the mass fraction of DM particles belonging to the GRASP in the torus, $\kappa$, which is defined as the mass of DM particles in the torus belonging to the GRASP divided by the total mass of DM particles in the torus. This fraction can range from 0.6\% to 33\%. These agree with the fractions estimated in ref.~\cite{Evans:2018bqy} for the DM particles belonging to the GRASP in the Solar neighbourhood. As noted in ref.~\cite{Evans:2018bqy}, the fraction of DM particles in the GRASP cannot be too high, since a large fraction could alter the observed MW rotation curves, change the ellipticity of the disc and the stellar kinematics, or modify the sphericity of the MW gravitational potential.

  \begin{table}[h!]
    \centering
    \begin{tabular}{|c|c|c|}
      \hline
       Halo Name  & $\kappa$ & $\beta$  \\
       \hline
       Au5 (MW-like) & 0.12 & 0.61 \\
       Au9 (MW-like) & 0.17 & 0.48 \\
       Au10 & 0.027 & 0.64\\
       Au15 & 0.22 & 0.56 \\
       Au17  & 0.032 & 0.66\\
       Au18 & 0.037 & 0.73 \\
       Au22 (MW-like) & 0.0058 & 0.82 \\
       Au24 (MW-like) & 0.089  & 0.50 \\
       Au26 & 0.33 & 0.32 \\
       Au27 & 0.20 & 0.39 \\
      \hline
    \end{tabular}
\caption{Fraction, $\kappa$, and anisotropy parameter, $\beta$, of DM belonging to the GRASP in the torus region at the Solar circle for all Auriga halos containing the GRASP. The MW-like halos with the GRASP are also specified.}
    \label{tab:sausageprop}
  \end{table}

Next, we compute the anisotropy of the DM particles in the GRASP. The anisotropy parameter of a distribution function is defined as
\begin{equation}
\beta = 1 - \frac{\sigma_\theta^2 + \sigma_\phi^2}{2 \sigma_r^2},
\label{eq:beta}
\end{equation}
where $\sigma_\theta$, $\sigma_\phi$, and $\sigma_r$ are the tangential, azimuthal, and radial velocity dispersions, respectively. When $\sigma_\theta = \sigma_\phi = \sigma_r$, the distribution function is isotropic, and $\beta=0$. When $\sigma_r^2 \gg \sigma_\theta^2 + \sigma_\phi^2$, the velocity distribution is radially anisotropic and $\beta \approx 1$, while for $\sigma_r^2 \ll \sigma_\theta^2 + \sigma_\phi^2$, the velocity distribution is tangentially anisotropic (i.e.~circular orbit) and $\beta \rightarrow -\infty$. In the third column of table \ref{tab:sausageprop}, we list the anisotropy parameters of the  DM belonging to the GRASP in the torus region for the 10 Auriga halos containing the GRASP.  The anisotropy parameter is in the range of $\beta = [0.32 - 0.82]$ depending on the halo. We have checked that these estimates for $\beta$ are robust with respect to increasing the volume of the torus at the Solar circle. For example, halo Au22 has only 69 DM particles belonging to the GRASP in the torus, and $\beta$ changes by only $\sim6$~\% if we consider a larger torus with $5\leq\rho \leq 11$~kpc and $|z|\leq 2$~kpc. 

One important observation is that $\beta$ for the DM particles belonging to the GRASP in the Solar circle is always smaller than the anisotropy of stars in the GRASP within 10 kpc from the Sun, $\beta_{\star} \sim 0.9$~\cite{Belokurov2019}. This is contrary to the assumption made in ref.~\cite{Evans:2018bqy}; i.e.~that the DM and stellar particles belonging to the GRASP have the same anisotropy. 

\begin{figure}[t!]
\begin{center}
  \includegraphics[width=0.6\textwidth]{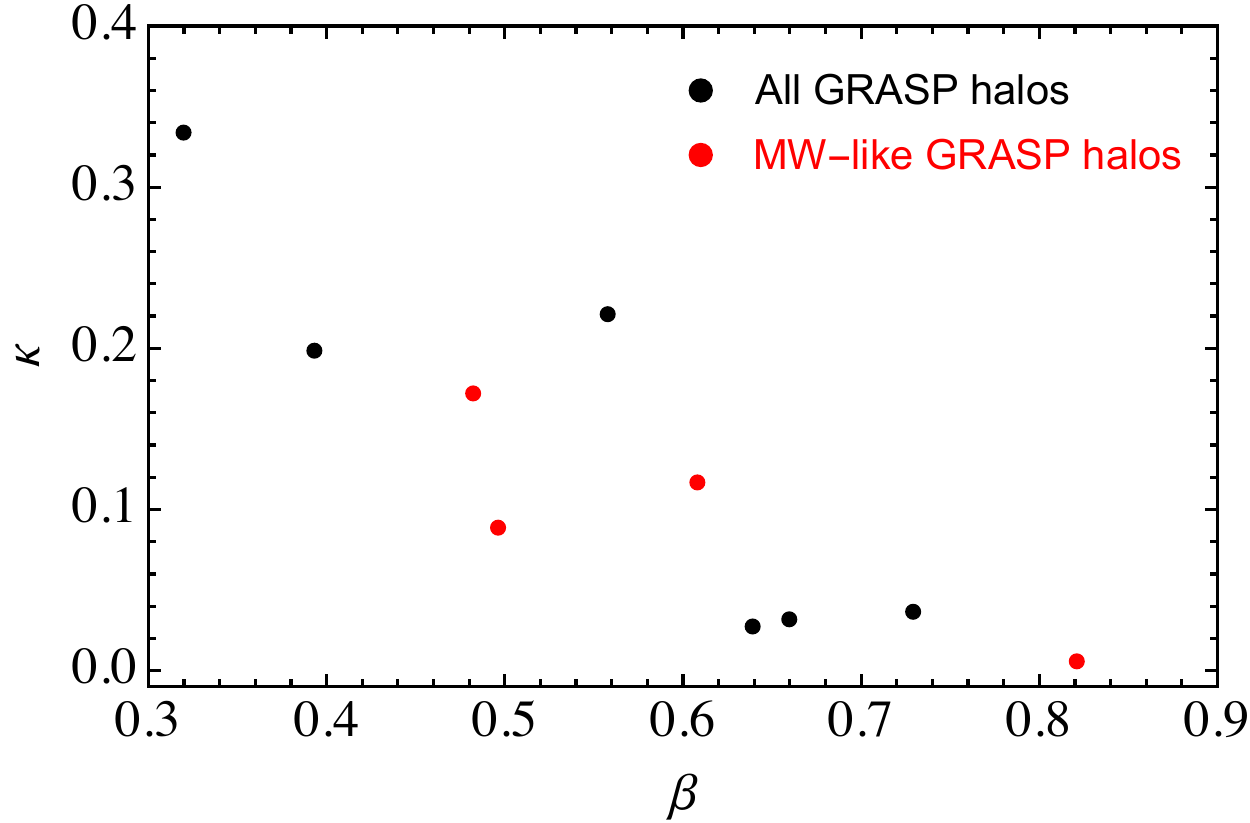}
\caption{Correlation between the fraction and anisotropy of the DM particles belonging to the GRASP in the torus. The red points specify the MW-like halos with the GRASP.}
\label{fig:kappabeta}
\end{center}
\end{figure}

\begin{figure}[t!]
\begin{center}
\includegraphics[width=0.48\textwidth]{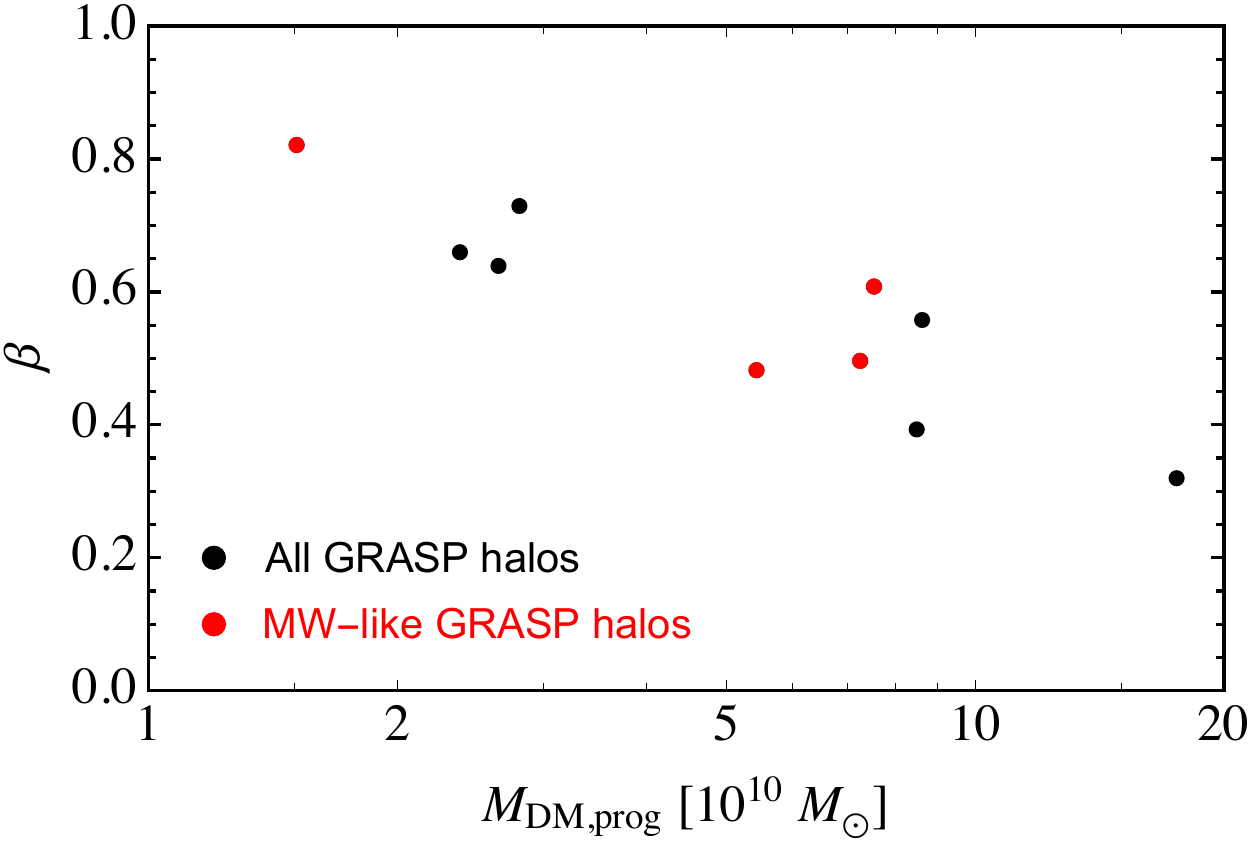}
\hspace{5pt}\includegraphics[width=0.48\textwidth]{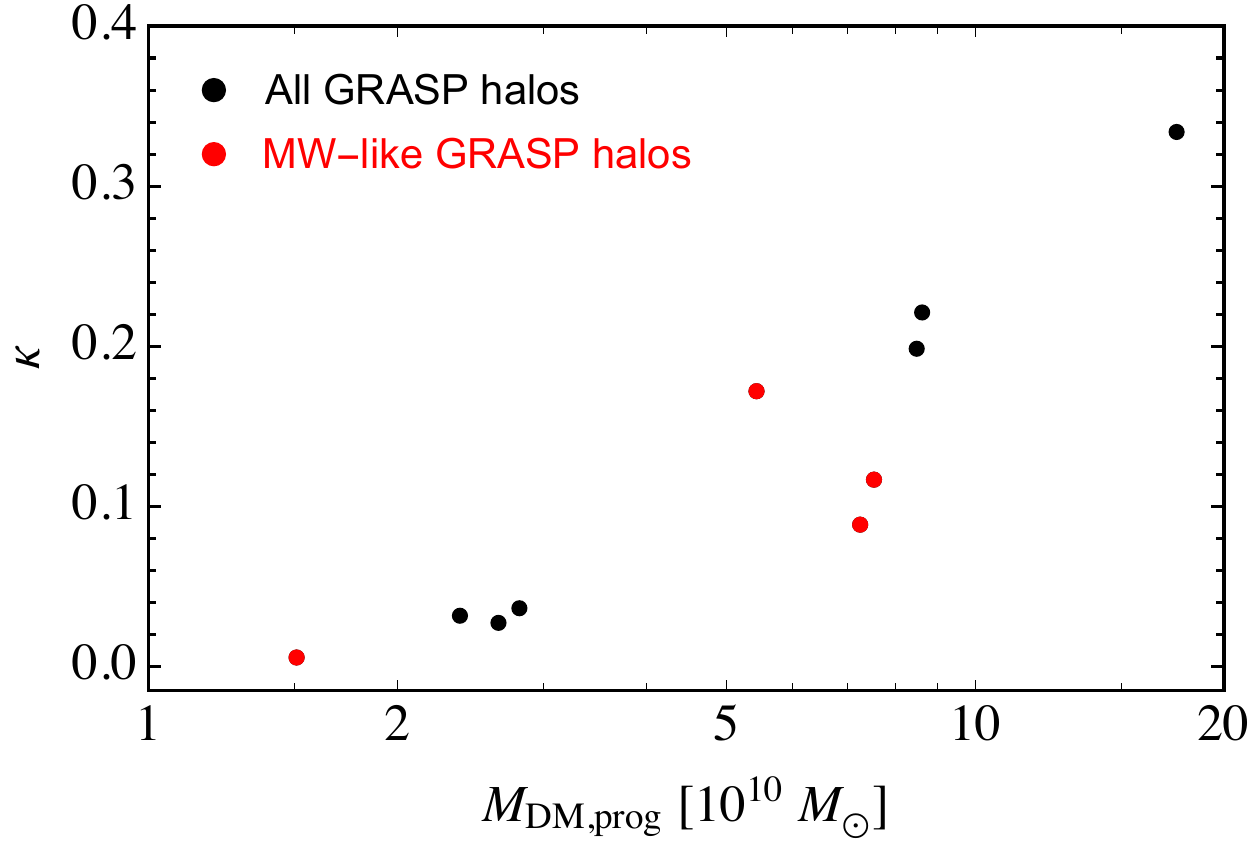}\\
\caption{Correlations between the anisotropy, $\beta$ (left panel), and the  fraction, $\kappa$ (right panel), of DM belonging to the GRASP in the torus region and the GRASP progenitor's DM mass, $M_{{\rm DM,prog}}$, at infall.}
\label{fig:kappaMDMStars}
\end{center}
\end{figure}

In figure~\ref{fig:kappabeta} we show that for the 10 Auriga halos with the GRASP, there exists an anti-correlation between the mass fraction, $\kappa$, and the anisotropy parameter, $\beta$, of the DM particles belonging to the GRASP in the torus. The MW-like halos are specified with red points in the figure. The larger the anisotropy parameter of the DM  particles of the GRASP, the smaller their fraction in the torus. The Pearson's correlation coefficient between $\kappa$ and $\beta$ is $-0.87$. This anti-correlation is a reflection of two other correlations displayed in   figure~\ref{fig:kappaMDMStars}. The left panel shows an anti-correlation between $\beta$ and the DM mass of the GRASP progenitor at infall, while the right panel shows a positive correlation between $\kappa$ and this mass. The correlation in the right panel is straightforward, but the origin of the correlation in the left panel is not as it results from an  interplay between tidal stripping, dynamical friction, and the dependence of the orbital parameters on subhalo mass at infall. This is the subject of an ongoing investigation, but for the purposes of the present paper it suffices to note the existence of the anti-correlation in figure~\ref{fig:kappabeta}.

\begin{figure}[t!]
\begin{center}
  \includegraphics[width=0.475\textwidth]{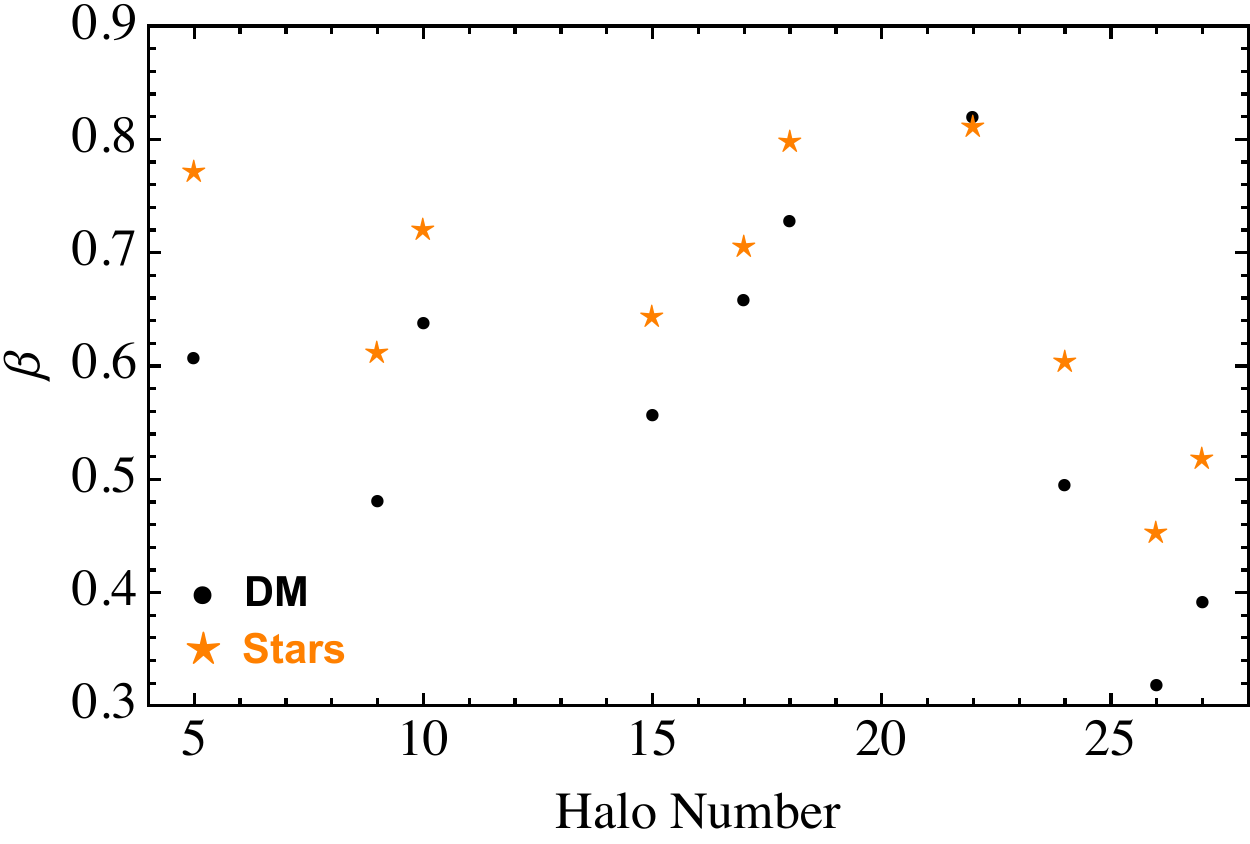}
\hspace{5pt}\includegraphics[width=0.50\textwidth]{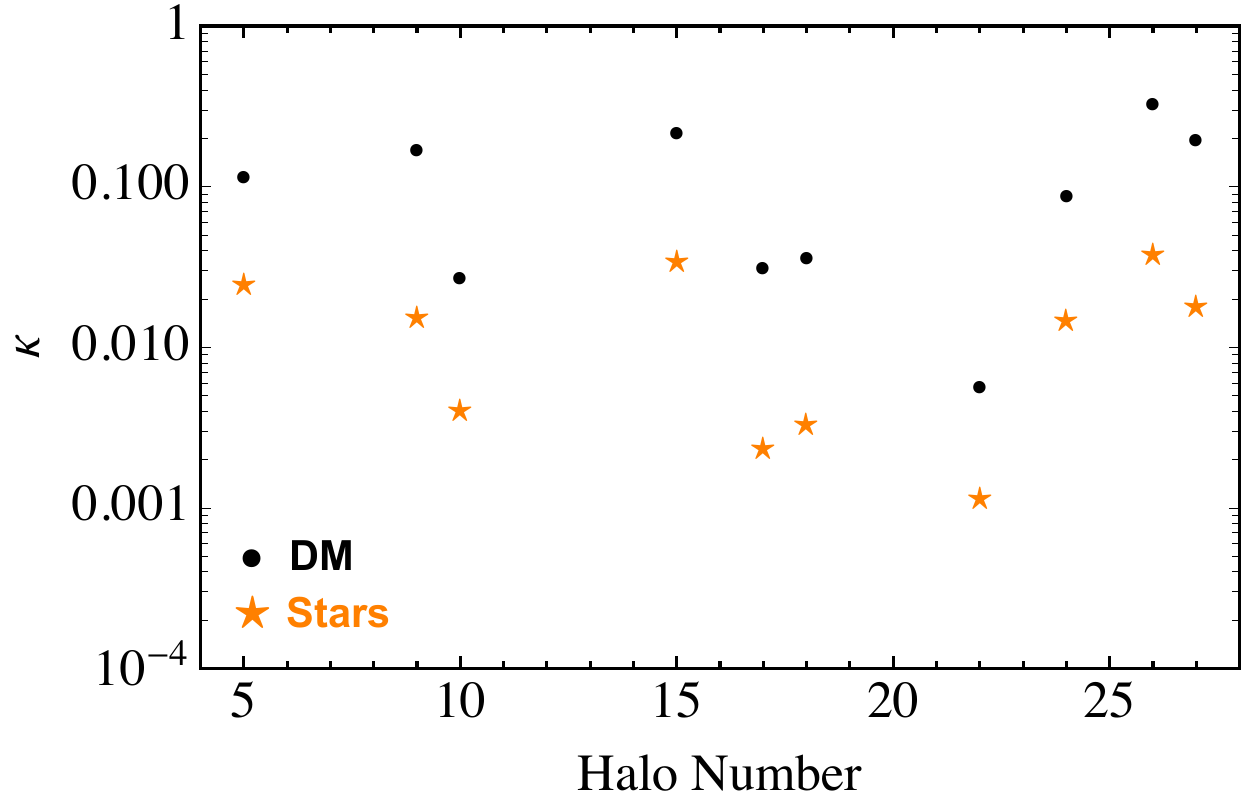}\\
\caption{Anisotropy parameter, $\beta$ (left panel), and mass fraction, $\kappa$ (right panel), of DM and stars belonging to the GRASP in the torus region for the 10 Auriga halos with the GRASP. The $x$-axis specifies the Auriga halo number.}
\label{fig:betakappaDMStars}
\end{center}
\end{figure}

Next, we compare the mass fraction and anisotropy of the DM and stars belonging to the GRASP in the torus region. The total number of star particles belonging to the GRASP in the torus region is in the range of $[60 - 4782]$ depending on the halo. For the four MW-like halos with the GRASP, the range is $[60 - 3400]$. As it can be seen from the left panel of figure~\ref{fig:betakappaDMStars}, the anisotropy of stars belonging to the GRASP in the torus is higher than the anisotropy of the DM brought in by the GRASP main progenitor in the same region. The difference is not surprising as DM and stars belonging to the GRASP progenitor experience different amounts of tidal stripping. Since DM particles are less concentrated than the stars in the progenitor, they tend to be stripped earlier, at larger distances from the centre, and thus from a different sector of the orbit compared to the stars. One possible reason for the systematic bias of stars relative to the DM in this panel, is that the orbits of massive subhalos tend to radialise with time~\cite{Amorisco2017}, so  stars that are stripped last tend to have higher eccentricities. From the right panel of figure \ref{fig:betakappaDMStars} one can see that the mass fraction of stars belonging to the GRASP in the torus region is lower than the mass fraction of DM from the same progenitor. This fraction is defined as the mass of stars or DM brought in by the GRASP main progenitor that ends up in the torus region divided by the total mass of stars or DM in the same volume. The stellar mass fractions that we find are significantly different from those reported in ref.~\cite{Fattahi2019} where  30\% to 50\% of the \emph{accreted} stars in the inner 20 kpc are found to originate from the GRASP progenitor. Here we do not distinguish between accreted and {\em in-situ} forming stars, and since the torus region lies on the Galactic plane the total mass in that region is dominated by in-situ stars.

%*******************************%

\section{Dark matter distribution}
\subsection{Dark matter density profiles}
\label{sec:density}

In this section, we present the DM density profiles for the Auriga MW-like halos with the GRASP and the local DM density for all Auriga MW-like halos.

\begin{figure}[t!]
\begin{center}
  \includegraphics[width=0.7\textwidth]{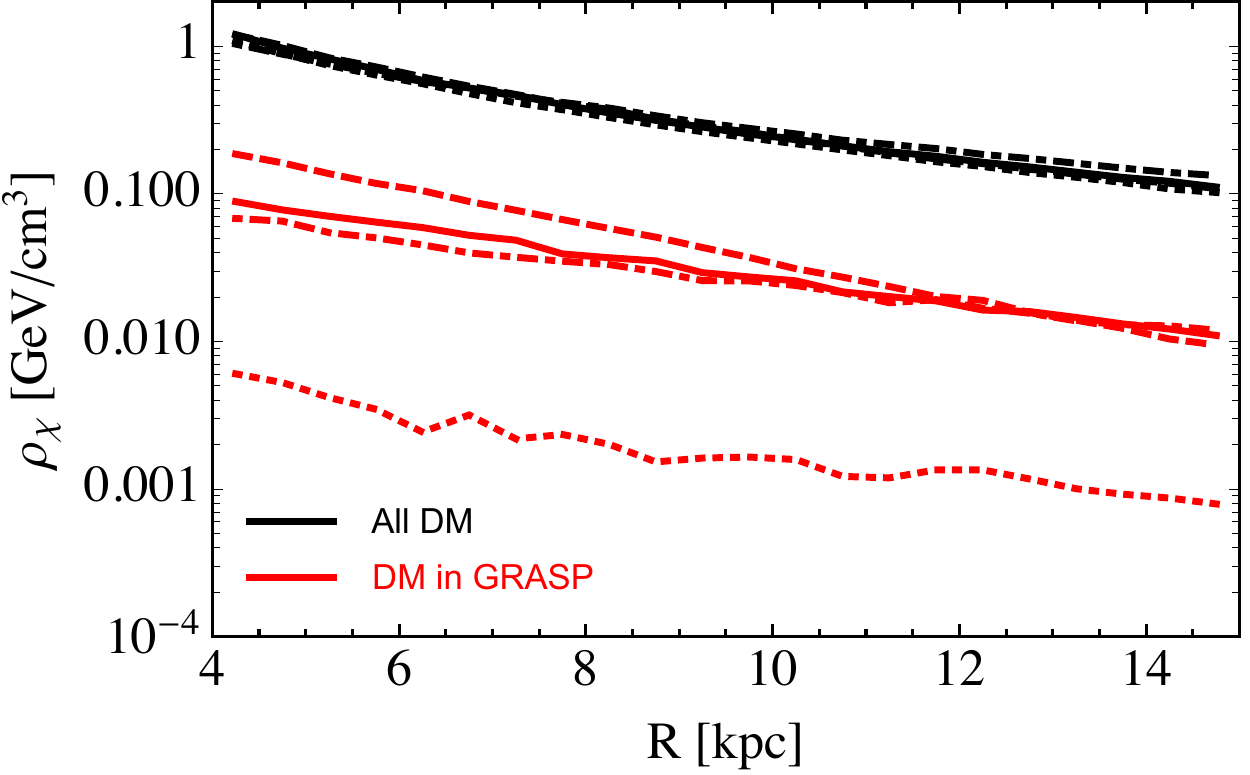}
\caption{DM density profiles for all DM (black) and DM belonging to the GRASP (red) for halos Au5 (solid), Au9 (dashed), Au22 (dotted), and Au24 (dot-dashed).}
\label{fig:rhoDM}
\end{center}
\end{figure}

Figure \ref{fig:rhoDM} shows the the average density profiles of all DM and the DM belonging to the GRASP, computed in spherical shells between $R=4-15$~kpc for the four Auriga MW-like halos with the GRASP. Clearly, the DM associated with the GRASP is a subdominant part of the total DM density in the inner halo. The total DM density profiles show very little variation between the four halos. This is not the case for the density profiles of DM in the GRASP which show a large halo-to-halo variation, with Au22 (dotted red curve in figure \ref{fig:rhoDM}) having the smallest density of DM in the GRASP, among the four halos. This is consistent with Au22 having the smallest fraction $\kappa$ of DM belonging to the GRASP in the torus region among all halos (see table \ref{tab:sausageprop}). 

Next, we compute the slope of the DM density profiles in the Solar neighbourhood, by fitting a power law $\rho_\chi(R) \propto R^{-a}$ to the density profiles of DM in the GRASP, DM not belonging to the GRASP, and all DM for $6 \leq R \leq 10$~kpc. The value of the slope, $a$, together with its standard error is given in table \ref{tab:slopes} for the four halos. One can see that the slope of the DM density profiles in the Solar neighbourhood is always less than $a=3$, and we do not find the  $R^{-3}$ dependence which was assumed in ref.~\cite{Evans:2018bqy}.

\begin{table}[h!]
    \centering
    \begin{tabular}{|c|c|c|c|}
      \hline
       Halo Name  & DM in GRASP  & DM not in GRASP & All DM\\
       \hline
       Au5 & $1.76 \pm 0.068$ & $1.94 \pm 0.023$  & $1.92 \pm 0.022$  \\
       Au9 & $2.30 \pm 0.055$ & $1.89 \pm 0.024$ & $1.95 \pm 0.022$  \\
        Au22 & $1.35 \pm 0.303$  & $1.89 \pm 0.023$ & $1.89 \pm 0.023$ \\
       Au24 & $1.29 \pm 0.075$  & $1.69 \pm 0.023$ & $1.66 \pm 0.022$ \\
      \hline
    \end{tabular}
    \caption{The slope and its standard error of the density profiles of DM in the GRASP, DM not belonging to the GRASP, and all DM in the Solar neighbourhood, $6 \leq R \leq 10$~kpc, for the MW-like Auriga halos with the GRASP.}
    \label{tab:slopes}
  \end{table}
  
  Direct detection event rates are proportional to the DM density in the Solar neighbourhood. In the SHM the fiducial value adopted for the local DM density is 0.3 GeV$/$cm$^3$. To extract the local DM density for the Auriga MW-like halos, we find the average DM density in the torus region in the Solar circle. We present the results in  Table~\ref{tab:Localrho}, which shows that for all simulated halos, the local DM density is larger than the fiducial value of 0.3~GeV$/$cm$^3$. This is similar to the results obtained from the EAGLE and APOSTLE hydrodynamic simulations for the local DM density~\cite{Bozorgnia:2016ogo}, and agrees with the global~\cite{McMillan:2011wd, Catena:2009mf, Weber:2009pt, Iocco:2011jz, Nesti:2013uwa, Sofue:2015xpa, Pato:2015dua, deSalas:2019pee} and local~\cite{Salucci:2010qr, Smith:2011fs, Bovy:2012tw, Garbari:2012ff, Zhang:2012rsb, Bovy:2013raa, 2018A&A...615A..99H, Buch:2018qdr} estimates from observations. Notice also that there is no significant difference between the values of the local DM density for halos with and without the GRASP.

  \begin{table}[t!]
    \centering
    \begin{tabular}{|c|c|}
      \hline
       Halo Name  &  $\rho_{\chi}^{\rm loc}$~[GeV$/$cm$^3$] \\
       \hline
       Au2 & 0.479 \\
       Au4 & 0.398 \\
       ~~~~~Au5 ($\star$) & 0.444 \\
       Au7 &  0.386  \\
       ~~~~~Au9 ($\star$) & 0.449 \\
       Au12 & 0.427 \\
       Au19 & 0.437 \\
       Au21 & 0.444 \\
      ~~~~~Au22 ($\star$)  & 0.370 \\
       ~~~~~Au24 ($\star$) & 0.483 \\
      \hline
    \end{tabular}
\caption{The average DM density in the torus region in the Solar circle for the Auriga MW-like halos. The halos with the GRASP are identified with a ($\star$) symbol.}
    \label{tab:Localrho}
  \end{table}

%*******************************%

\subsection{Dark matter velocity distributions}
\label{sec:velocities}

The DM velocity distribution in the Solar neighbourhood is an important input in the calculation of direct detection event rates. In this section we discuss and compare the local DM velocity distributions in the Galactic rest frame for the MW-like Auriga halos with and without the GRASP. Moreover, to study the effect of baryons on the local DM distribution, we compare the velocity distributions of the hydrodynamic halos with their DMO counterparts. 

We extract the local DM speed distribution by computing the  average speed distribution of DM particles in the torus region in the Solar circle. For the DMO halos, since there is no Galactic disc to align the torus with, we instead orient the torus in the same direction of the torus in their hydrodynamic halo counterparts. The speed distribution, $f(v)$, is normalised as $\int dv f(v)=1$, and is related to the velocity distribution, $\tilde{f}({\bf v})$, by 
\begin{equation}
f(v) = v^2 \int d\Omega_{\bf v} \tilde{f}({\bf v}),
\end{equation}
where $d\Omega_{\bf v}$ is an infinitesimal solid angle around the direction ${\bf v}$, and $\int d^3 v \tilde{f}({\bf v}) =1$.

\begin{figure}[t!]
 \begin{center}
   \includegraphics[width=0.48\textwidth]{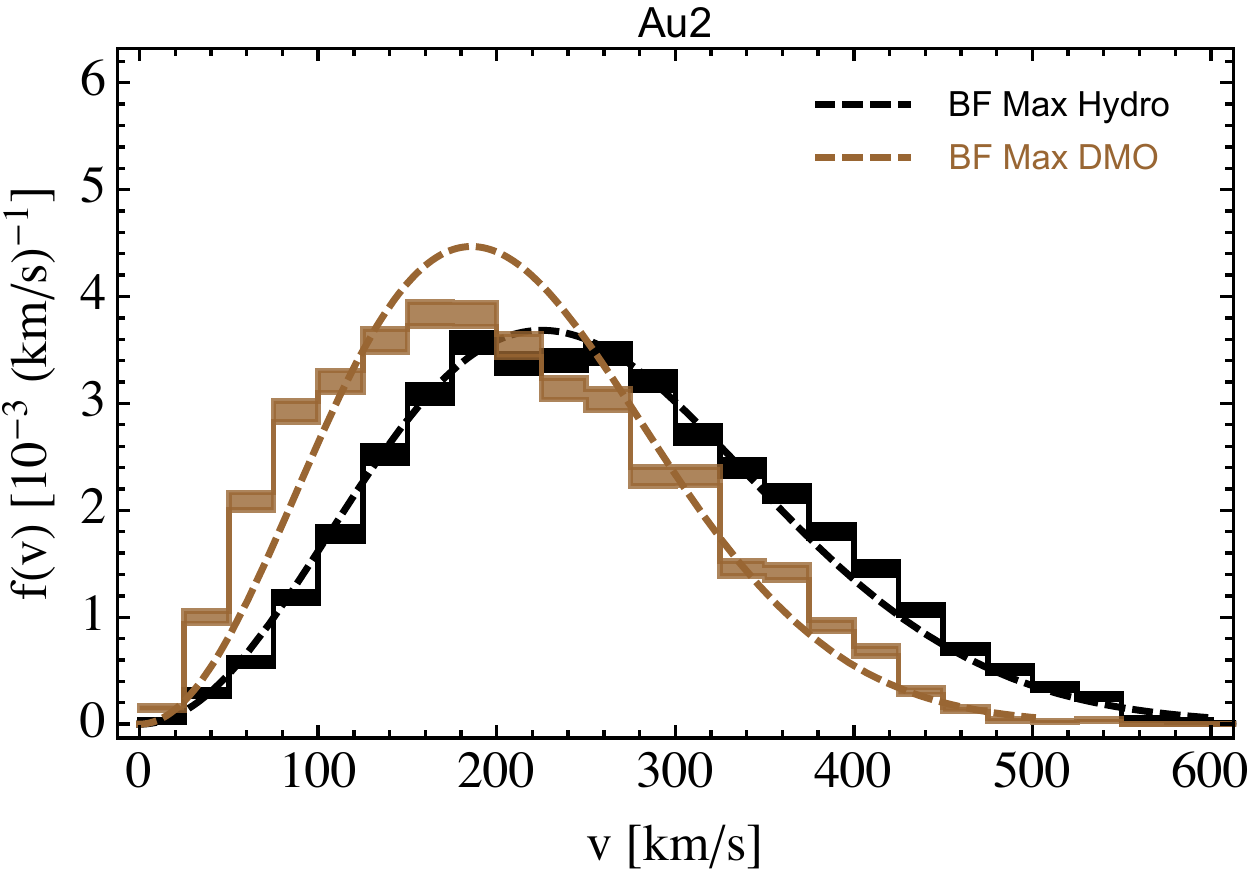}
   \includegraphics[width=0.48\textwidth]{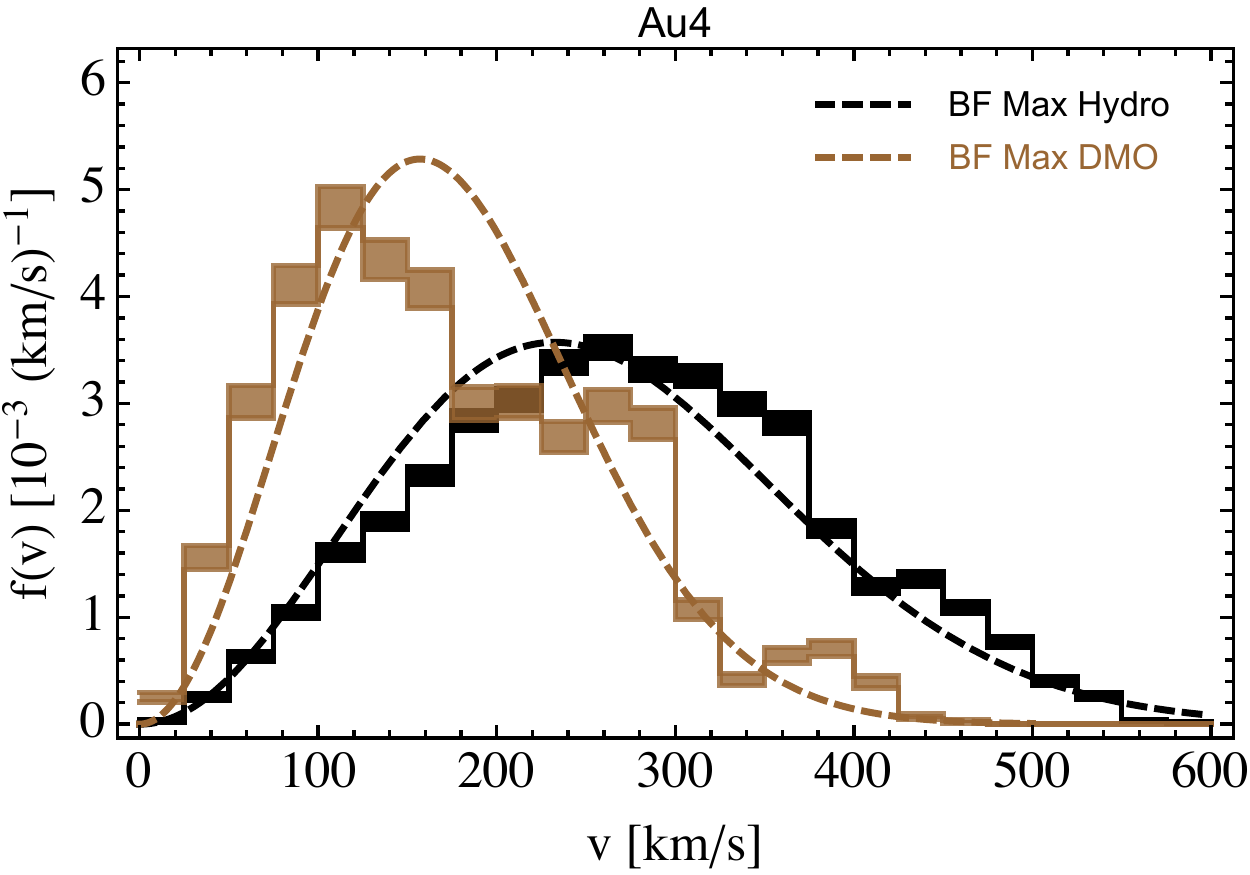}\\
   \includegraphics[width=0.48\textwidth]{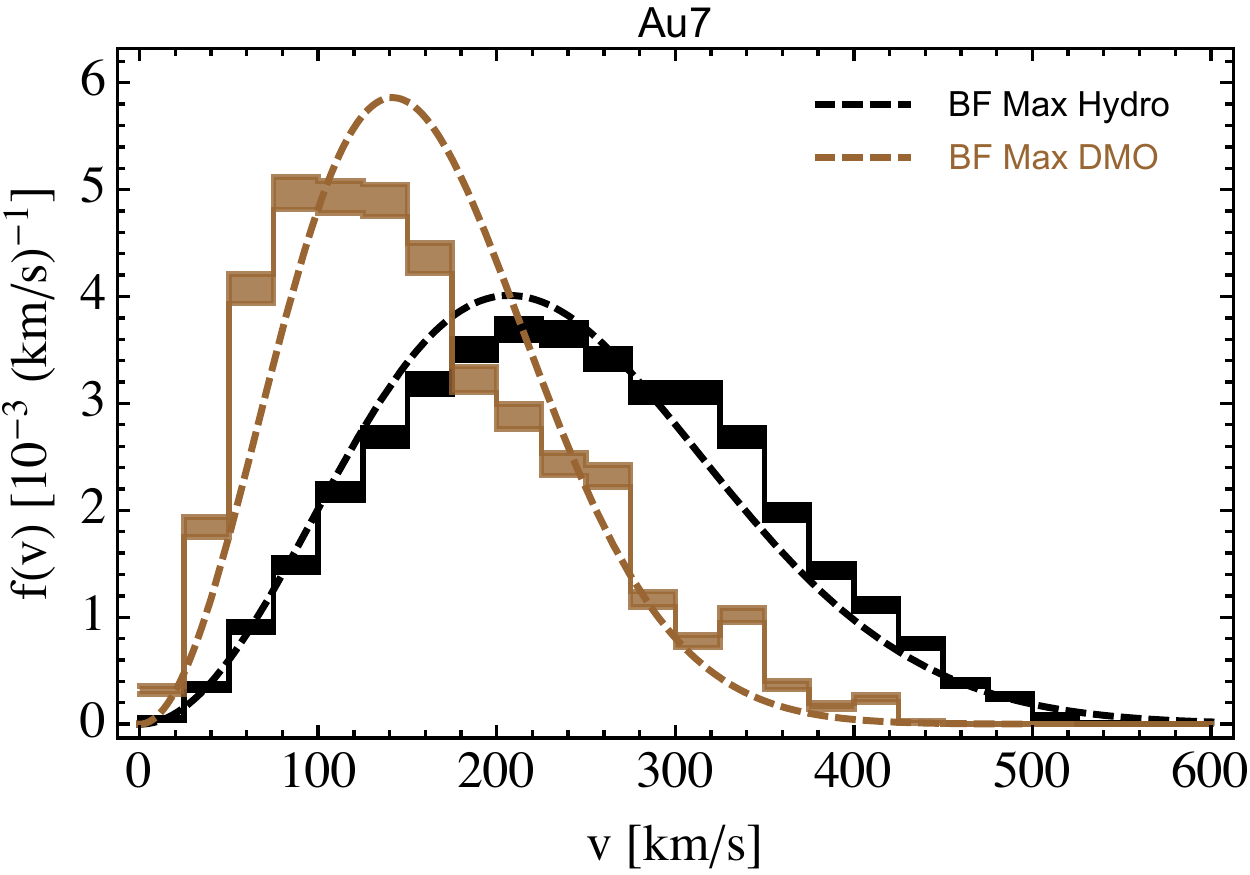}
   \includegraphics[width=0.48\textwidth]{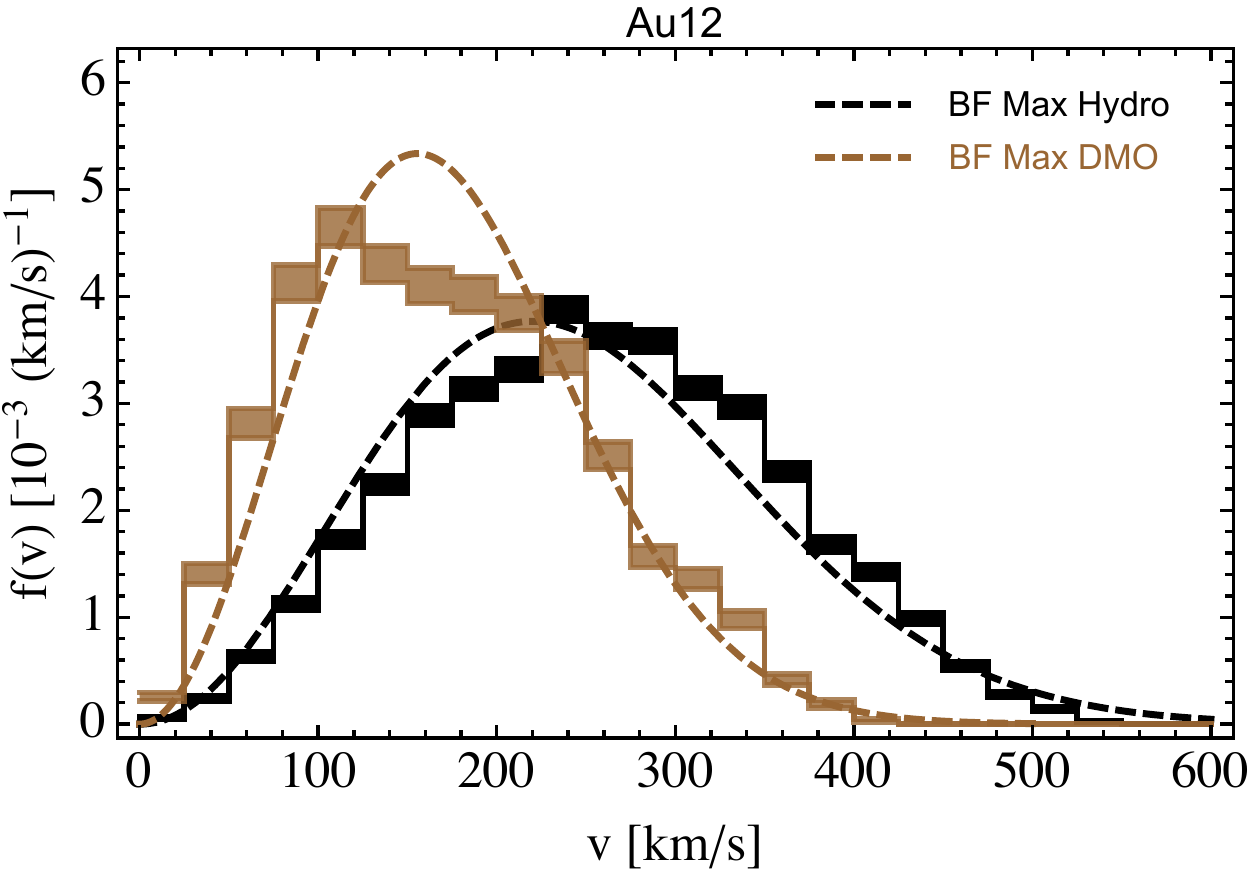}\\
   \includegraphics[width=0.48\textwidth]{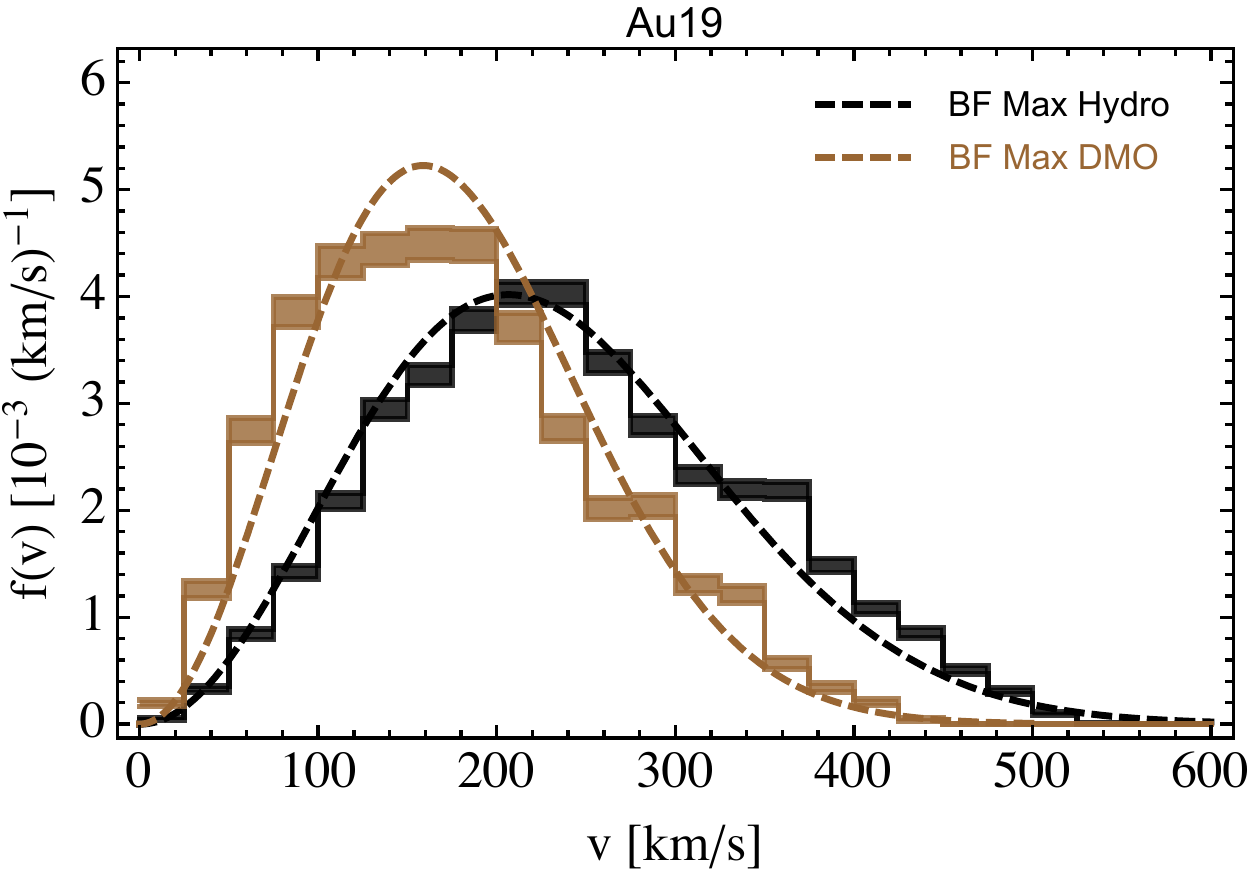}
   \includegraphics[width=0.48\textwidth]{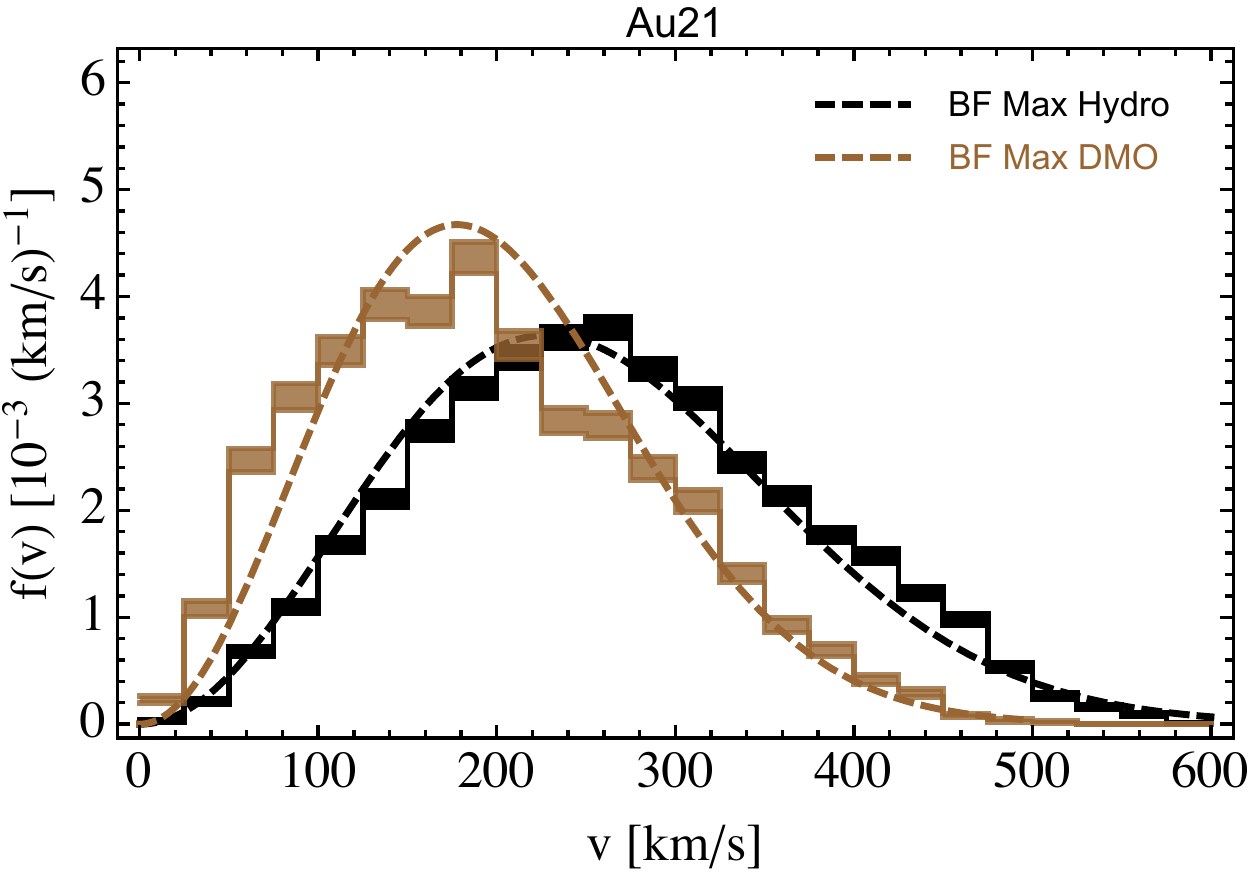}\\
 \caption{Local DM speed distributions in the Galactic rest frame (shaded bands specifying the $1\sigma$ Poisson errors) for the hydrodynamic Auriga MW-like halos without the GRASP (black shaded bands) and their DMO counterparts (brown shaded bands). The dashed black and brown curves are the best fit Maxwellian (BF Max) distributions for the hydrodynamic and DMO cases, respectively.}
 \label{fig:fv}
 \end{center}
 \end{figure}

\begin{figure}[t!]
 \begin{center}
 \includegraphics[width=0.48\textwidth]{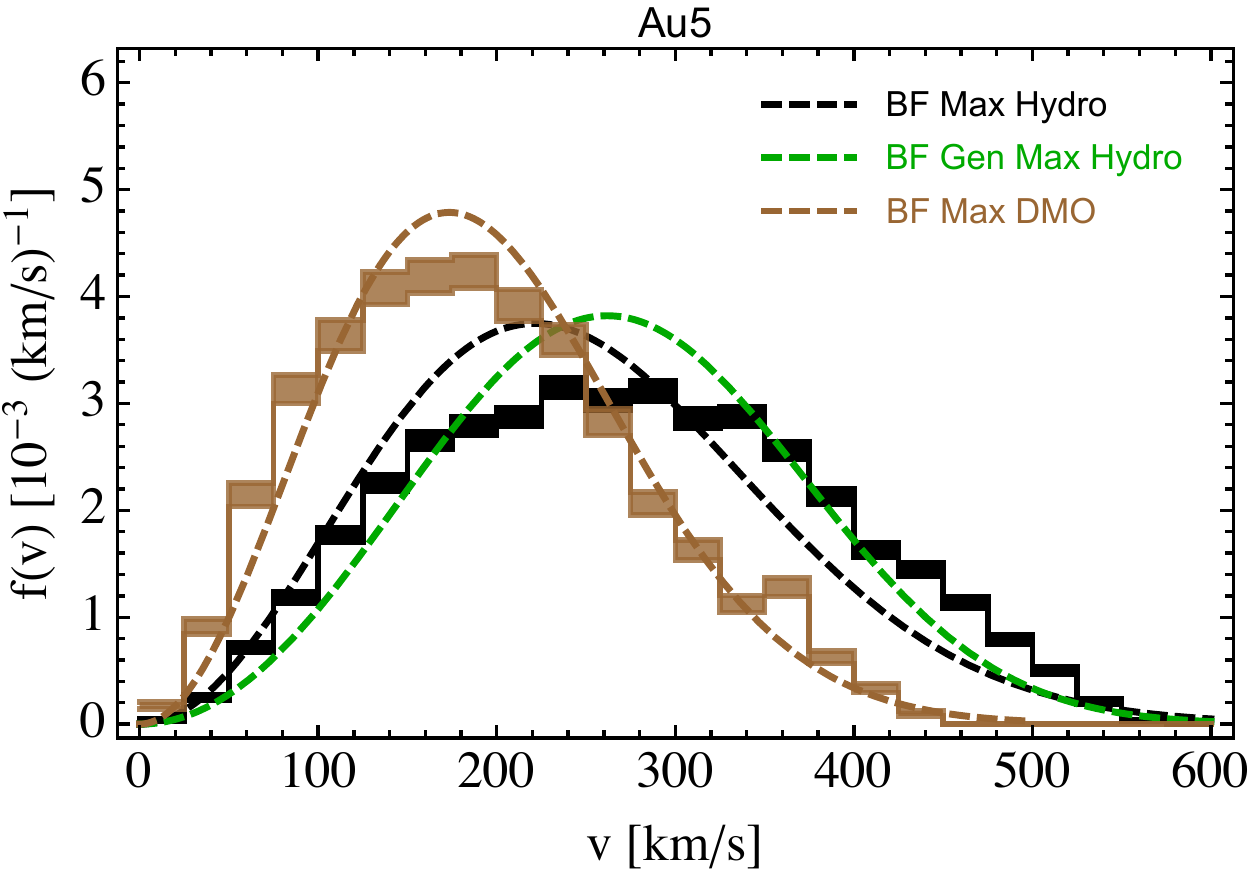}
   \includegraphics[width=0.48\textwidth]{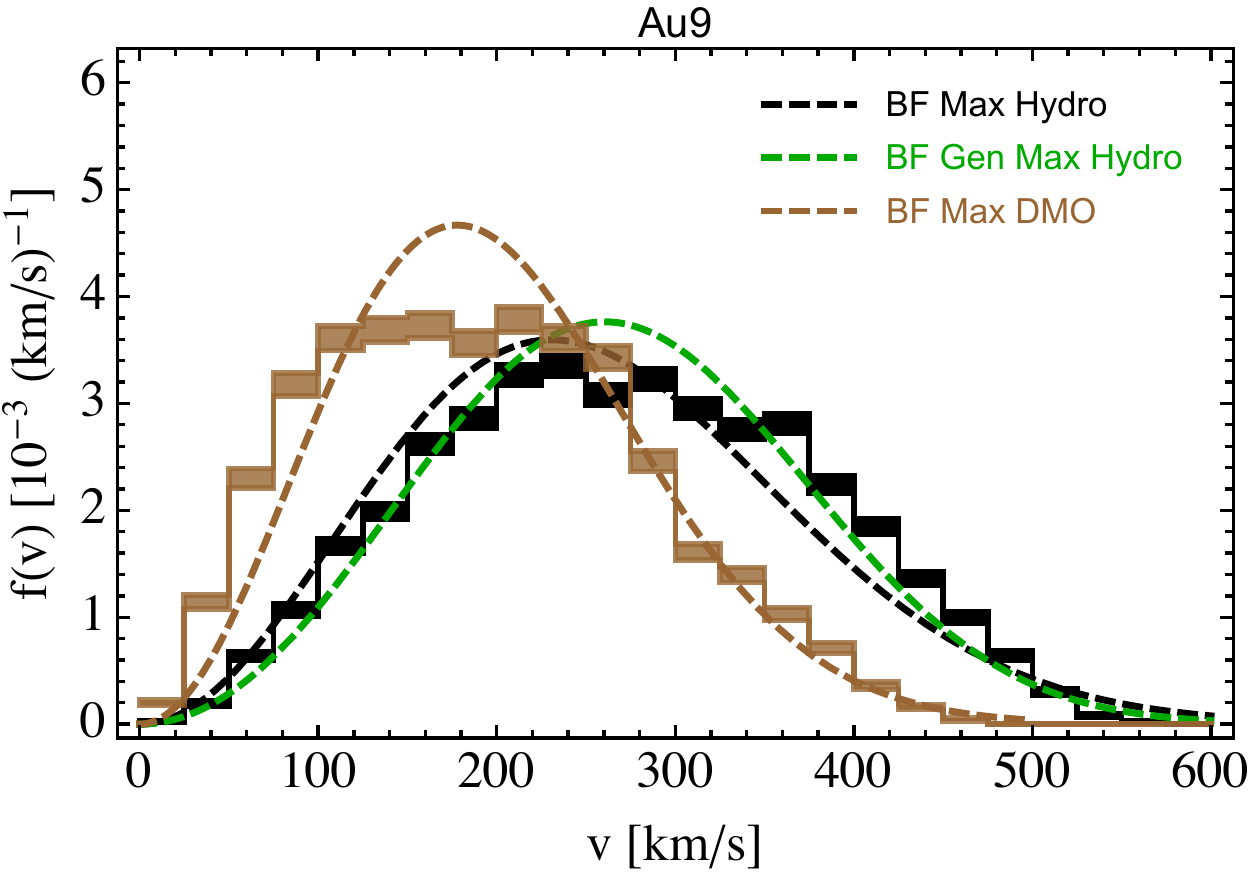}\\
   \includegraphics[width=0.48\textwidth]{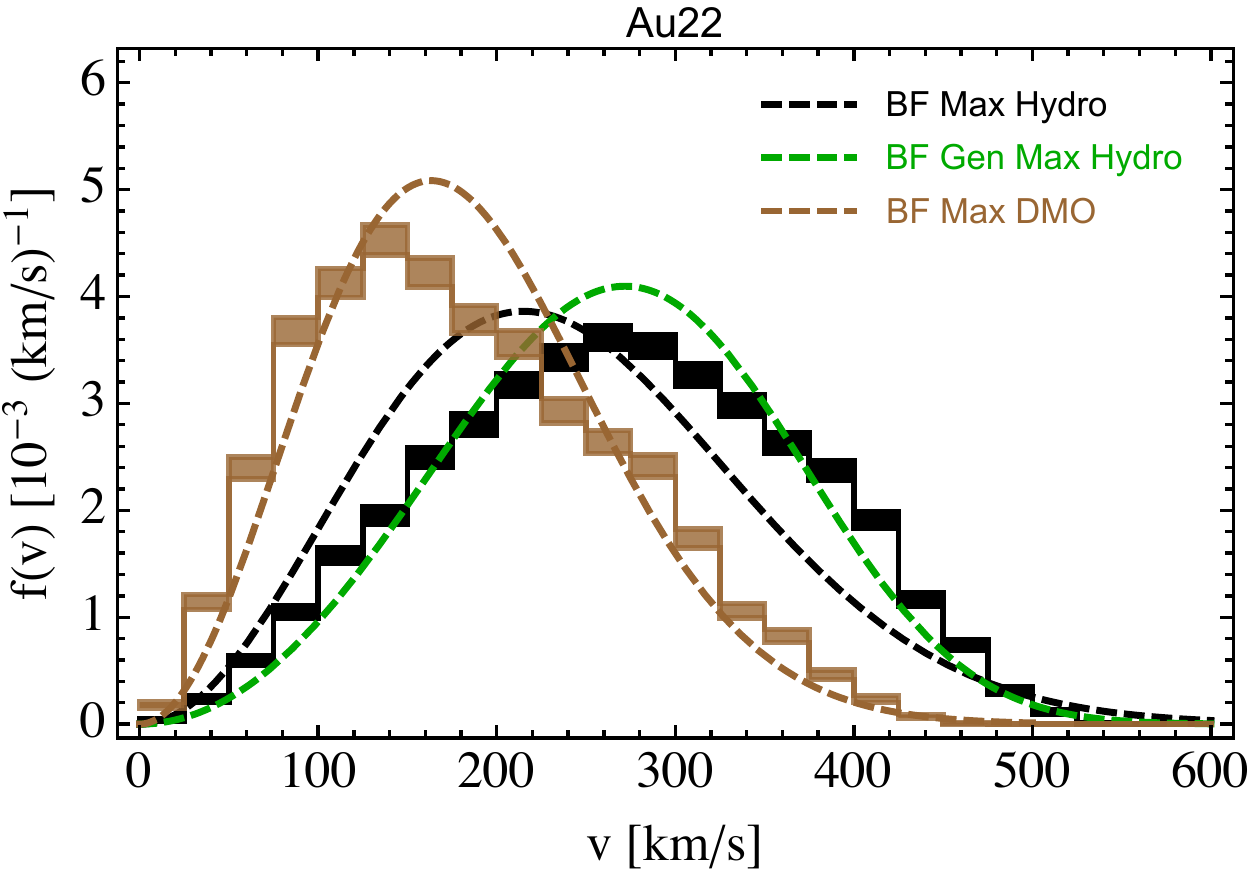}
   \includegraphics[width=0.48\textwidth]{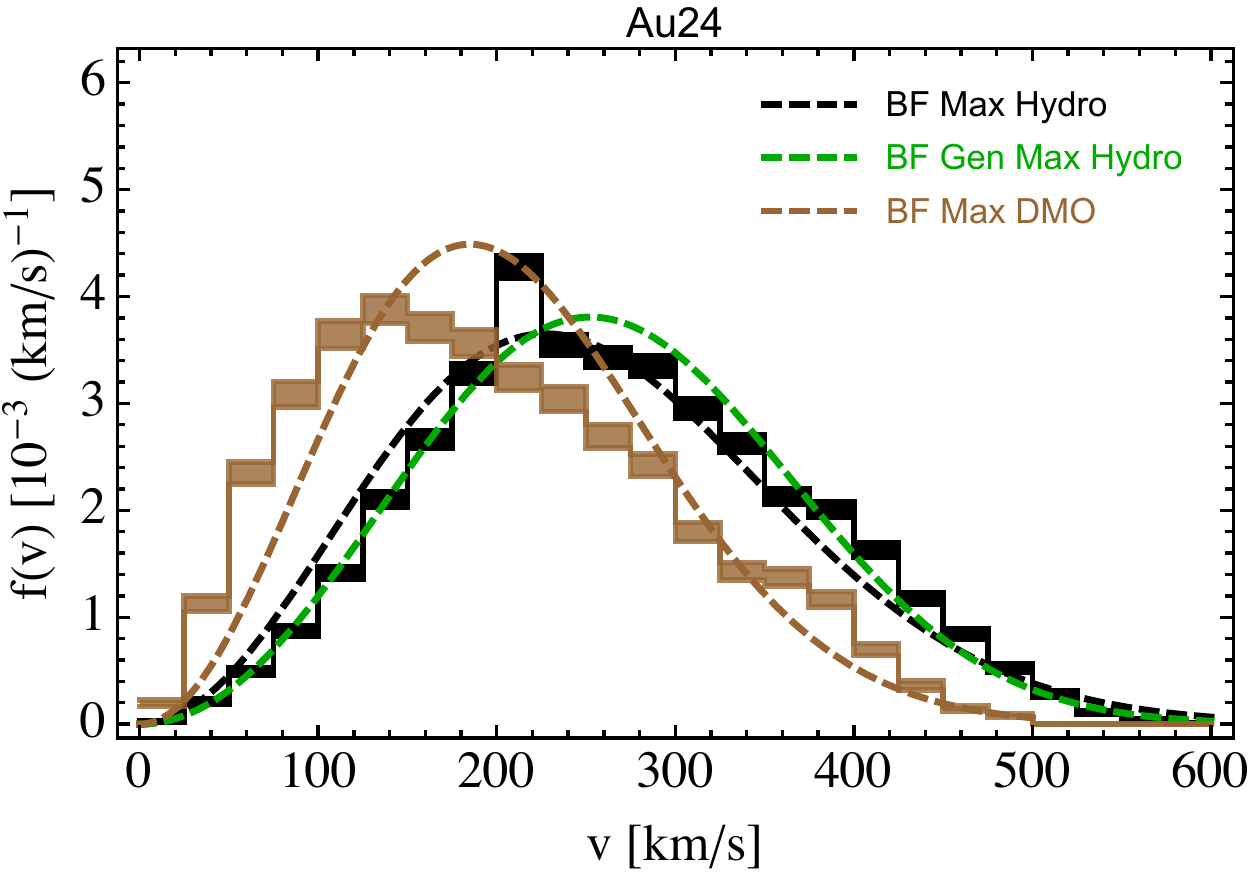}
 \caption{Same as figure~\ref{fig:fv}, but for the halos with the GRASP. The dashed green curves are the best fit generalized Maxwellian distributions for the hydrodynamic halos.}
 \label{fig:fv-GRASP}
 \end{center}
 \end{figure}

In figures \ref{fig:fv} and \ref{fig:fv-GRASP} we present the local DM speed distributions in the Galactic rest frame for the MW-like Auriga halos without and with the GRASP, respectively. The shaded black (brown) curves specify the $1\sigma$ Poisson error band for the hydrodynamic (DMO) halos. The speed bin size is 25~km~s$^{-1}$, which is the optimal size found to avoid having large fluctuations in the data, as well as being small enough to reveal the features in the distributions. The dashed black (brown) curves show the best fit Maxwellian speed distributions for each hydrodynamic (DMO) halo.

As expected, we can see from figures \ref{fig:fv} and \ref{fig:fv-GRASP} that including baryons in the simulations increases the peak speed of the DM speed distributions. This feature was also observed in other recent simulations~\cite{Bozorgnia:2016ogo, Bozorgnia:2017brl, Kelso:2016qqj, Sloane:2016kyi}. This is due to baryons making the MW gravitational potential deeper, which results in higher DM speeds in the Solar neighbourhood.

\begin{figure}[t!]
\begin{center}
  \includegraphics[width=0.48\textwidth]{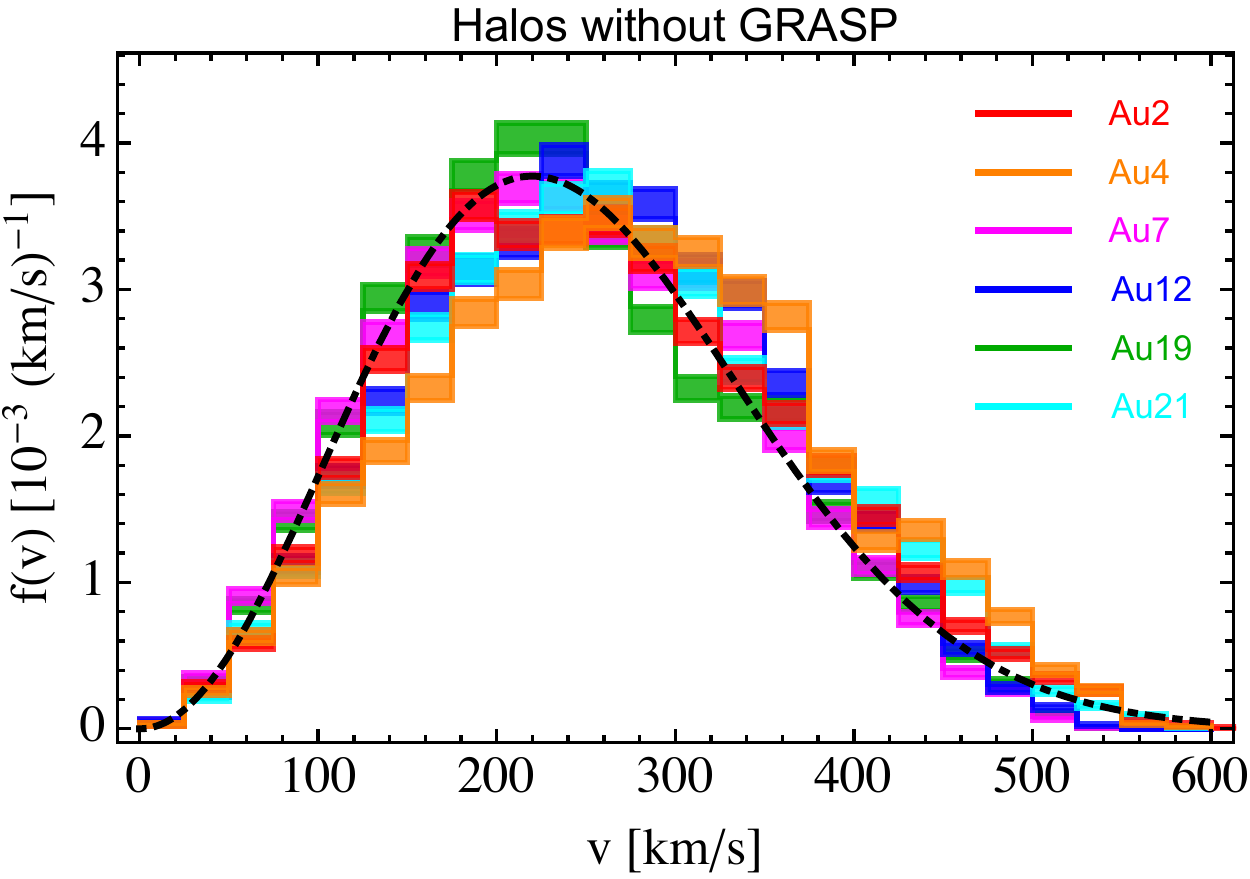}
  \includegraphics[width=0.48\textwidth]{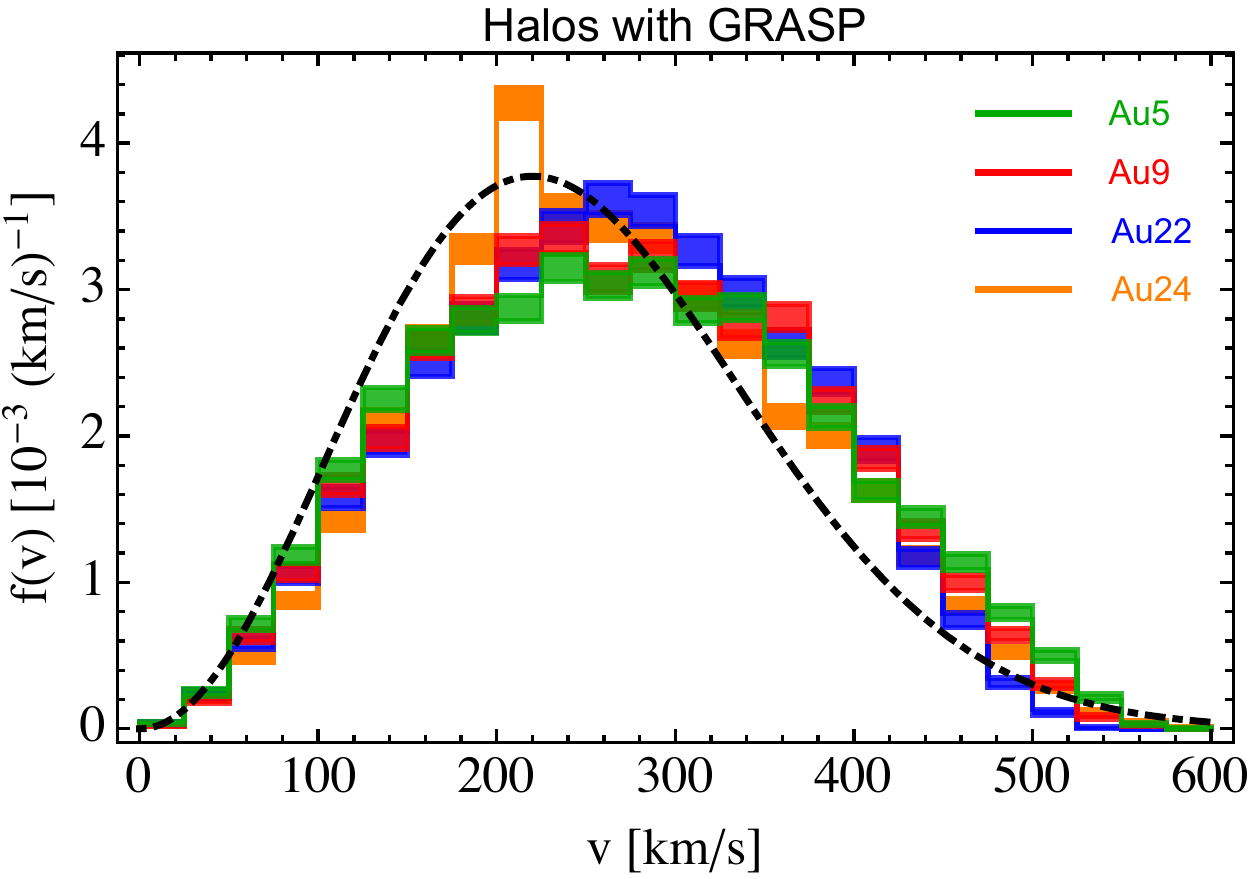}
\caption{A comparison of the local DM speed distributions in the Galactic rest frame for the Auriga MW-like halos without (left panel) and with (right panel) the GRASP (shaded colour bands), and the SHM Maxwellian speed distribution with a peak speed of 220~km~s$^{-1}$ (black dot-dashed curve). }
\label{fig:fvGRASP-SHM}
\end{center}
\end{figure}

In figure~\ref{fig:fvGRASP-SHM} we show a comparison of the local DM speed distributions of the MW-like hydrodynamic halos without (left panel) and with (right panel) the GRASP (same as the black shaded bands shown  in figures~\ref{fig:fv} and~\ref{fig:fv-GRASP}, respectively), and the SHM Maxwellian speed distribution with a peak speed of 220~km~s$^{-1}$. The speed distributions of the halos with the GRASP are all shifted to higher speeds compared to the SHM Maxwellian. Since the MW-like halos without the GRASP are more numerous, their DM speed distributions show a larger halo-to-halo variation compared to the halos with the GRASP.

The Galactic escape speeds\footnote{The escape speeds are calculated using spherically averaged potential estimates, and are defined as the minimum speed required to reach $3 R_{200}$, where $R_{200}$ is the virial radius.} at 7 kpc (inner radius of our torus region) are in the range of $v_{\rm esc}=[531 - 605]$~km~s$^{-1}$ for the MW-like simulated halos without the GRASP, and $v_{\rm esc}=[537 - 582]$~km~s$^{-1}$ for the halos with the GRASP. These values agree well with the estimates for the MW escape speed from observations~\cite{Piffl:2013mla, Piffl:2014mfa, 2018A&A...616L...9M, 2019MNRAS.485.3514D}.

Next, we discuss how well the local DM speed distributions of the MW-like halos with and without the GRASP fit different distribution functions. To perform the fit of the various distribution functions to the DM speed distributions of the simulated halos, we minimise the $\chi^2$ function,
\begin{equation}
\chi^2({\bf p}) = \sum_{i=1}^N \frac{(y_i - f(v_i, {\bf p}))^2}{\sigma_i^2}.
\end{equation}
Here, ${\bf p}$ specifies the free parameters in each distribution function with their best fit values determined by the $\chi^2$ minimisation; $y_i$ is the value of the speed distribution at speed $v_i$, with $\sigma_i$ its $1\sigma$ Poisson error, and $N$ is the  number of bins in speed for evaluating the speed distributions.

First, we consider the standard Maxwellian distribution, $f(v) \propto v^2 \exp(-v^2/v_0^2)$, with the peak speed, $v_0$, as a free parameter. In table~\ref{tab:BFMax} we list the best fit peak speeds and the reduced $\chi^2$ (i.e.~$\chi^2_{\rm red}$) values for the goodness of fit to the local DM speed distributions of all MW-like halos. The fit to the Maxwellian distribution is in general not good for any of the halos, but the halos with the GRASP show worse fits, as it can be seen from their high $\chi^2_{\rm red}$ values. In the last two columns of table~\ref{tab:BFMax},  the best fit peak speeds and $\chi^2_{\rm red}$ for the DMO counterparts of the Auriga MW-like halos are given. For four out of the six halos without the GRASP, the fit to the Maxwellian distribution is worse for the DMO compared to the hydrodynamic halo. This is similar to the conclusions of previous simulations, showing that in most cases baryons make the DM distribution more Maxwellian~\cite{Bozorgnia:2016ogo, Bozorgnia:2017brl, Kelso:2016qqj, Sloane:2016kyi}. In halos Au12 and Au19, the fit is slightly better for the DMO halo compared to the hydrodynamic case. Furthermore, in most of the hydrodynamic halos with the GRASP, the $\chi^2_{\rm red}$ for the fit to the Maxwellian is larger than their DMO counterparts. This is expected since the DMO counterparts of the GRASP halos do not contain baryons or the GRASP component, and hence their local DM speed distributions are expected to be smoother than their hydrodynamic counterparts with the GRASP. However, the best fit Maxwellian distributions in general do not fit the DMO halos well either.

Next, we consider the generalized Maxwellian distribution given by,
\begin{equation} 
\label{eq:genMax}
f(v) \propto  v^2  \exp[ -( v / v_0 )^{2 \alpha} ] \, ,
\end{equation}
with free parameters $v_0$ and $\alpha$. Notice that the standard Maxwellian
distribution is obtained by setting $\alpha = 1$. In table~\ref{tab:BFMax} we list the best fit parameters and the reduced $\chi^2$ for the fit of the local DM speed distribution of the hydrodynamic MW-like halos to the generalized Maxwellian distribution. Due to the additional parameter in the generalized Maxwellian, for all halos the fit is better compared to the standard Maxwellian fit. In figure~\ref{fig:fv-GRASP}, the dashed green curves specify the best fit generalized Maxwellian speed distribution for each halo with the GRASP. Although the fit to the DM speed distributions of halos with the GRASP is highly improved for the generalized Maxwellian, it is not a very good fit yet, as it can be seen from the $\chi^2_{\rm red}$ values in table~\ref{tab:BFMax}.

 \begin{table}[t!]
    \centering
    \begin{tabular}{|c|c|c|c|c|c||c|c|c|}
      \hline
       & \multicolumn{2}{|c|}{Maxwellian} & \multicolumn{3}{|c||}{Generalized Maxwellian} 
       &  \multicolumn{2}{|c|}{Maxwellian DMO}\\
      \hline
       Halo Name  & $v_0$~[km~s$^{-1}$] & $\chi^2_{\rm red}$ & $v_0$~[km~s$^{-1}$] & $\alpha$ & $\chi^2_{\rm red}$ & $v_0$~[km~s$^{-1}$] & $\chi^2_{\rm red}$\\
       \hline
       Au2 & 225.39 & 11.63 & 248.18 & 1.12 & 8.45 & 185.83 & 26.91\\
       Au4 & 232.45 & 30.38 & 292.58 & 1.36 & 13.49 & 157.06 & 34.15\\
      ~~~~~Au5 ($\star$) & 221.51 & 65.19 &  296.44 & 1.42 & 37.00 & 173.45 & 11.22\\
       Au7 & 207.06 & 27.19 & 250.16 & 1.26 & 15.19 & 141.68 & 50.23 \\
       ~~~~~Au9 ($\star$) & 230.96 & 34.96 & 292.05 & 1.37 & 18.37 & 177.88 & 28.85 \\
       Au12 & 220.31 & 30.24 & 281.28 & 1.41 & 7.98 & 155.56 & 17.41 \\
       Au19 & 206.67 & 23.94 & 235.27 & 1.17 & 18.16 & 158.88 & 19.04\\
       Au21 & 228.24 & 16.73 & 267.79 & 1.22 &  7.86 & 177.69 & 19.27\\
       ~~~~~Au22 ($\star$) & 215.00 & 83.17 & 319.39 & 1.77 & 19.61 & 163.31 & 26.19 \\
      ~~~~~Au24 ($\star$) & 227.43 & 31.29 & 280.07 & 1.32 & 10.24 & 184.95 & 33.48 \\
      \hline
    \end{tabular}
\caption{Best fit parameters and the reduced $\chi^2$ values for the goodness of fit of the Maxwellian  and generalized Maxwellian distributions to the local DM speed distributions of the Auriga MW-like halos. Halos with the GRASP are specified with a ($\star$). For comparison, the best fit peak speed and $\chi^2_{\rm red}$ of the Maxwellian distribution for the DMO halos are given in the last two columns.}
\label{tab:BFMax}
 \end{table}

We next check if the bimodal DM distribution considered in ref.~\cite{Evans:2018bqy} provides a better fit to the local DM speed distributions of the halos containing the GRASP. In particular, ref.~\cite{Evans:2018bqy} considers a  linear combination of a Maxwellian and an anisotropic velocity distribution, given by,
\begin{equation} 
\label{eq:GRASPfv}
\tilde{f}({\bf v}) = (1-\kappa) \tilde{f}_{\rm Max}({\bf v}) + \kappa \tilde{f}_{\rm Anis}({\bf v}),
\end{equation}
where $\tilde{f}_{\rm Max}({\bf v})$ is the standard Maxwellian distribution, and
\begin{equation}
\tilde{f}_{\rm Anis}({\bf v}) \propto \exp\left(-\frac{v_r^2}{2\sigma_r^2}-\frac{v_\theta^2}{2\sigma_\theta^2}-\frac{v_\phi^2}{2\sigma_\phi^2}\right).
\end{equation}
Here $v_r$, $v_\theta$, and $v_\phi$ are the radial, tangential, and azimuthal DM velocities, respectively. For each halo with the GRASP, we  fix the fraction, $\kappa$, and anisotropy parameter, $\beta$ of the DM belonging to the GRASP in the torus region to their actual values derived for each halo given in table~\ref{tab:sausageprop}. Notice that  $\sigma_r$, $\sigma_\theta$, and $\sigma_\phi$ are functions of the peak speed of the anisotropic distribution, $v_0^{\rm Anis}$, and the anisotropy parameter, $\beta$. By fixing $\kappa$ and $\beta$, there remains two free parameters in eq.~\eqref{eq:GRASPfv}, which are the peak speeds of the Maxwellian, $v_0^{\rm Max}$, and the anisotropic distributions, $v_0^{\rm Anis}$. We  fit the speed distribution derived from eq.~\eqref{eq:GRASPfv} to the local DM speed distributions of the  halos with the GRASP. The best fit parameters and the $\chi^2_{\rm red}$ for the goodness of fits are given in table~\ref{tab:BFSaus} for the four  halos with the GRASP. We find that even with the additional free parameter, the combination of the Maxwellian and anisotropic velocity distributions does not improve the fit to the local DM speed distributions of the halos with the GRASP compared to the standard Maxwellian. This is most likely due to the anti-correlation observed between $\kappa$ and $\beta$ (see figure~\ref{fig:kappabeta}). Namely, for halos with a large $\beta$ for the GRASP DM particles in the torus, the fraction of those DM particles is so small that they do not play a significant role in the total local DM speed distribution. On the other hand, for  halos where the fraction of DM belonging to the GRASP in the torus is larger (e.g.~reaching 17\% for halo Au9), their anisotropy parameter is not significant, minimising the effect of the GRASP DM component on the total local DM speed distribution.

 \begin{table}[h!]
    \centering
    \begin{tabular}{|c|c|c|c|}
      \hline
       & \multicolumn{3}{|c|}{Maxwellian + Anisotropic}\\ 
      \hline
       Halo Name  & $v_0^{\rm Max}$~[km~s$^{-1}$] &  $v_0^{\rm Anis}$~[km~s$^{-1}$] & $\chi^2_{\rm red}$ \\
       \hline
      Au5 & 222.09 & 209.31 & 73.22 \\
      Au9 & 231.06 & 230.02 & 38.79 \\
      Au22 & 215.05 & 194.05 & 88.04 \\
      Au24 & 227.46 & 226.05 & 34.31 \\
      \hline
    \end{tabular}
\caption{Same as table~\ref{tab:BFMax}, but for the fit of the local DM speed distribution of the GRASP halos to the distribution function given in eq.~\eqref{eq:GRASPfv}. The best fit peak speeds of the Maxwellian and the anisotropic distributions are specified as $v_0^{\rm Max}$ and $v_0^{\rm Anis}$, respectively}
\label{tab:BFSaus}
 \end{table}
 
Finally, for completeness we consider the RICE distribution function, given by,
\begin{equation}
f(v) \propto v \exp{\left[\frac{-(v^2 + \alpha^2)}{2v_0^2}\right]}I_0\left(\frac{\alpha v}{v_0^2}\right),
\end{equation}
where $I_0(x)$ is the modified Bessel function of the first kind with order zero, and $v_0$ and $\alpha$ are the two free parameters. We find that the RICE distribution provides a better fit to the local DM speed distributions of the halos with the GRASP compared to the standard Maxwellian distribution, but  a worse fit compared to the generalized Maxwellian distribution. Hence, we do not present here the best fit parameters of the RICE distribution.

\begin{figure}[t!]
\begin{center}
  \includegraphics[width=0.6\textwidth]{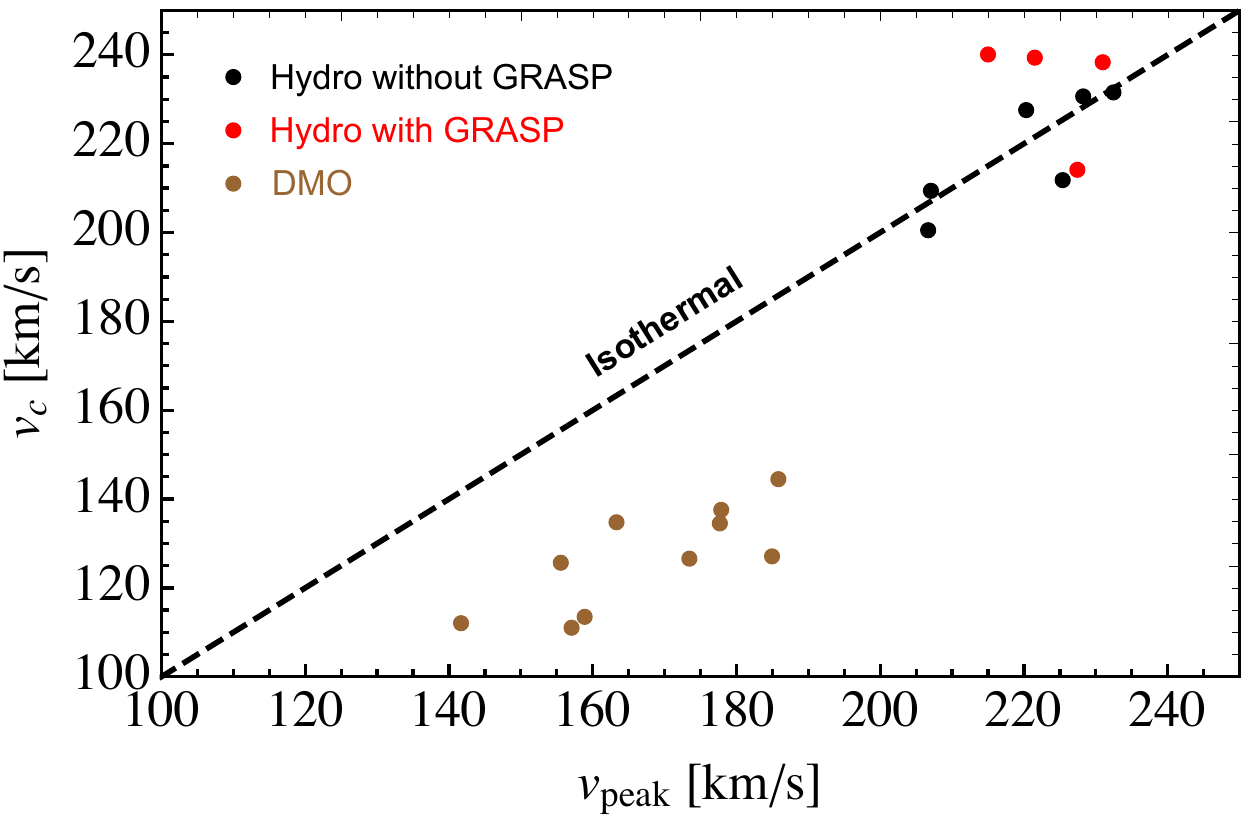}
\caption{A comparison of the local circular speed, $v_c$, at 8 kpc, and the best fit Maxwellian peak speed, $v_{\rm peak}$, for the fit to the local DM speed distributions of the hydrodynamic MW-like halos without the GRASP (black data points), with the GRASP (red data points), and the DMO halos (brown data points). The case of an isothermal halo with $v_{\rm peak}=v_c$ is shown as a dashed black line.}
\label{fig:vpeakvc}
\end{center}
\end{figure}

To understand how the local DM speed distributions of the MW-like halos compare to an isothermal distribution, in figure \ref{fig:vpeakvc} we show a comparison of the best fit peak speeds of the Maxwellian distribution, $v_{\rm peak}$, and the local circular speeds, $v_c$, for each hydrodynamic and DMO halo. The local circular speeds are computed at $R=8$~kpc from the total enclosed mass within a sphere with a radius of 8 kpc for each halo. In the case of the isothermal halo, $v_c = v_{\rm peak}$. Comparing the hydrodynamic MW-like halos with their DMO counterparts, we can see that baryons make the local halos closer to isothermal. This is expected, and similar to the results obtained for the EAGLE simulations in ref.~\cite{Bozorgnia:2017brl, Bozorgnia:2016ogo}. Due to the lack of baryons in the DMO simulations, the rotation curves of the DMO halos are still rising at the Solar position, and do not reach a plateau until larger radii. For the hydrodynamic halos, however, the rotation curves are close to flat at $R=8$~kpc (see figure \ref{fig:vc}), and the halos are hence closer to isothermal at the Solar circle. Notice that the hydrodynamic halos with the GRASP are slightly further from  isothermal compared to the halos without the GRASP. However, one point which should be emphasised here is that although the best fit Maxwellian peak speeds of the hydrodynamic halos without the GRASP are close to their local circular speeds, as discussed before, their fit to the Maxwellian distribution is in general not good (see table~\ref{tab:BFMax}).

We do not test or discuss other distribution functions for fitting the local DM speed distributions of the simulated halos in this work. The relevant quantity for computing direct detection event rates for standard interactions of DM with nuclei, is not directly the DM speed distribution, but an integrated quantity, called the \emph{halo integral} which we will discuss in section~\ref{sec:halintegral}. In the next section, we will present the halo integrals obtained from the MW-like Auriga halos, and evaluate how they compare to the standard Maxwellian and generalized Maxwellian halo integrals.

%*******************************%

\section{Implications for direct dark matter detection}
\label{sec:dirdet}

In this section, we discuss the implications of the DM distributions extracted from the Auriga MW-like halos with and without the GRASP for the interpretation of the results of direct DM detection experiments. 

\subsection{Halo integrals}
\label{sec:halintegral}

Direct DM detection experiments aim to measure the small recoil energy of a target nucleus in an underground detector after a scattering with a DM particle $\chi$. The differential event rate is given by
\begin{equation}
\frac{dR}{dE_R} = \frac{\rho_\chi^{\rm loc}}{m_\chi}\frac{1}{m_T} \int_{v>v_{\rm min}} d^3 v ~\frac{d\sigma_T}{d E_R}~v~\tilde{f}_{\rm det}({\bf v}, t),
\label{eq:difrate}
\end{equation}
where $E_R$ is the recoil energy of the target nucleus, $m_\chi$ and $m_T$ are the masses of the DM and target nucleus, respectively, $d\sigma_T/dE_R$ is the differential DM-nucleus scattering cross section, $\rho_\chi^{\rm loc}$ is the local DM density, $\tilde{f}_{\rm det}({\bf v}, t)$ is the local DM velocity distribution in the detector rest frame, and ${\bf v}$ is the relative velocity between the DM and the nucleus, with $v \equiv |{\bf v}|$. 

Assuming the DM-nucleus scattering is elastic, the minimum speed required for the DM particle to deposit a recoil energy $E_R$ in the detector is given by,
\begin{equation}
v_{\rm min} = \sqrt{\frac{m_T E_R}{2 \mu^2_{\chi T}}},
\end{equation}
where $\mu_{\chi T}$ is the DM-nucleus reduced mass. 

To study the implications of the DM component of the GRASP for the interpretation of direct detection results, a very useful quantity to analyse is the DM \emph{halo integral} which along with $\rho_\chi^{\rm loc}$ contains the astrophysical dependence of the event rate. For the case of standard spin-independent and spin-dependent DM-nucleus interactions, the differential event rate becomes proportional to the halo integral,
\begin{equation}
\eta(v_{\rm min}, t) \equiv \int_{v>v_{\rm min}} d^3 v ~ \frac{\tilde{f}_{\rm det}({\bf v}, t)}{v}.
\label{eq:halointegral}
\end{equation}

To extract the halo integrals of the simulated halos, we boost the local DM velocity distributions of the halos from the Galactic frame to the detector reference frame, by $\tilde{f}_{\rm det}({\bf v}, t) = \tilde{f}_{\rm gal}({\bf v}+{\bf v}_s+{\bf v}_e(t))$. Here ${\bf v}_e(t)$ is the velocity of the Earth with respect to the Sun, and ${\bf v}_s = {\bf v}_c + {\bf v}_{\rm pec}$ is the velocity of the Sun in the Galactic rest frame, where ${\bf v}_c$ is the Sun's circular velocity, and ${\bf v}_{\rm pec}=(11.10, 12.24, 7.25)$~km~s$^{-1}$~\cite{Schoenrich:2009bx} (in Galactic coordinates) is the peculiar velocity of the Sun with respect to the Local Standard of Rest. When boosting to the detector rest frame,  we take the local circular speed, $|{\bf v}_c|=v_c$, extracted from the mass enclosed within a sphere with radius 8 kpc for each halo. The values of $v_c$ for halos with and without the GRASP are shown in figure~\ref{fig:vpeakvc}. The time dependence in the halo integral originates from the velocity of the Earth with respect to the Sun, ${\bf v}_e(t)$. In the following, we discuss the \emph{time-averaged} halo integrals, i.e.~halo integrals averaged over one year.

\begin{figure}[t!]
\begin{center}
  \includegraphics[width=0.49\textwidth]{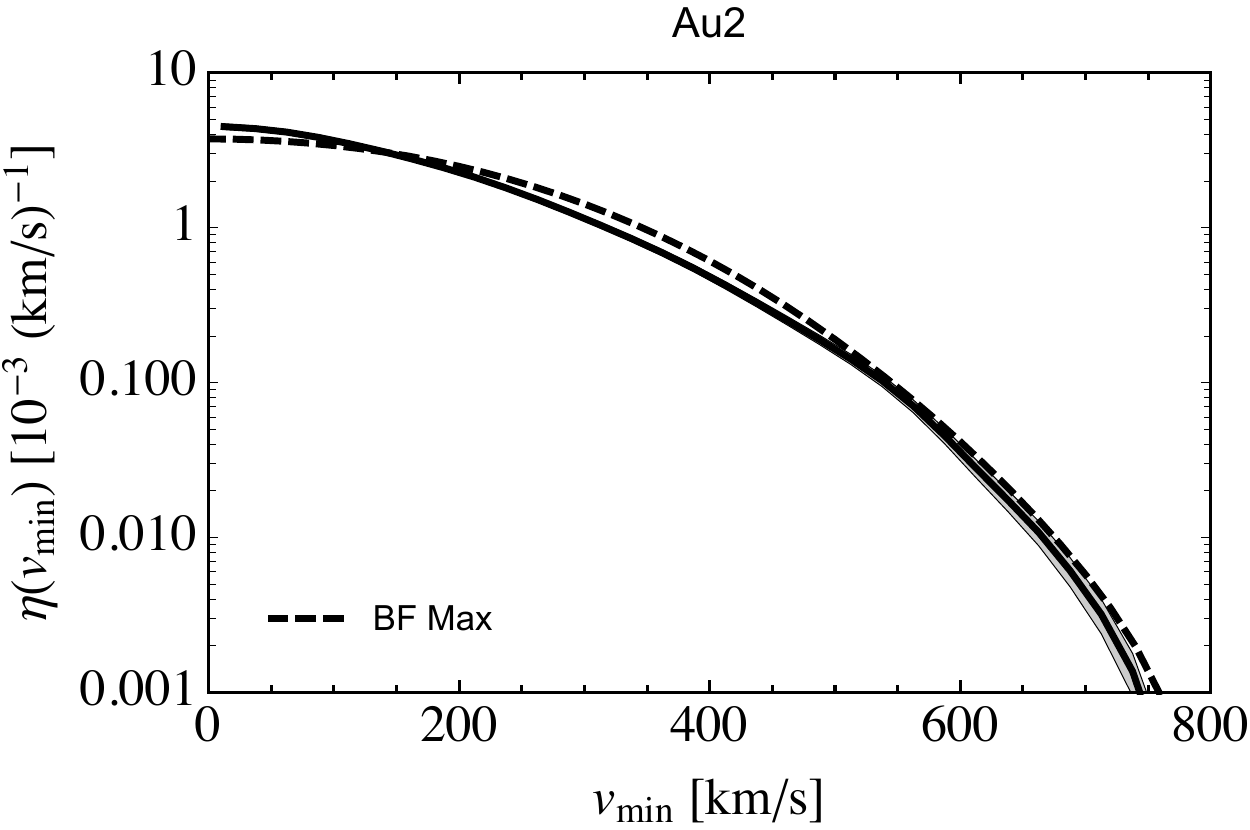}
  \includegraphics[width=0.49\textwidth]{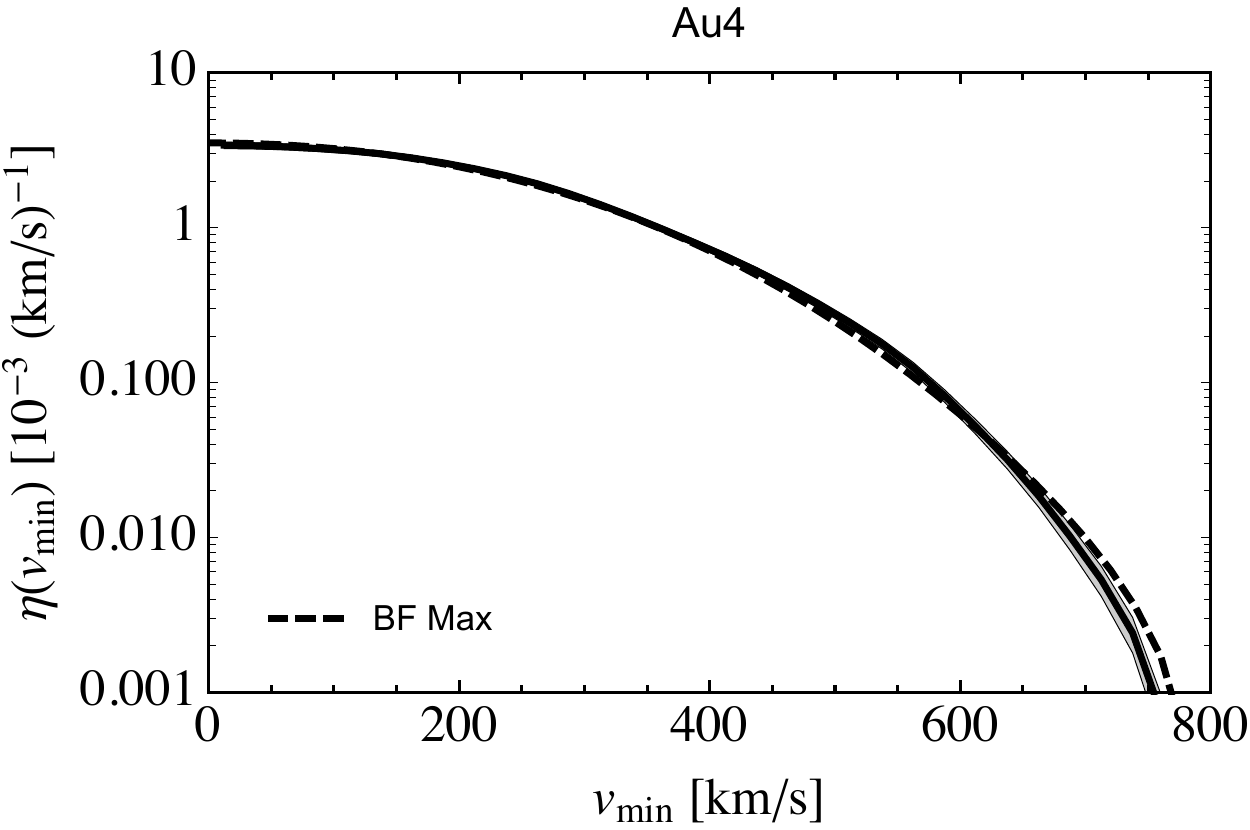}\\
  \includegraphics[width=0.49\textwidth]{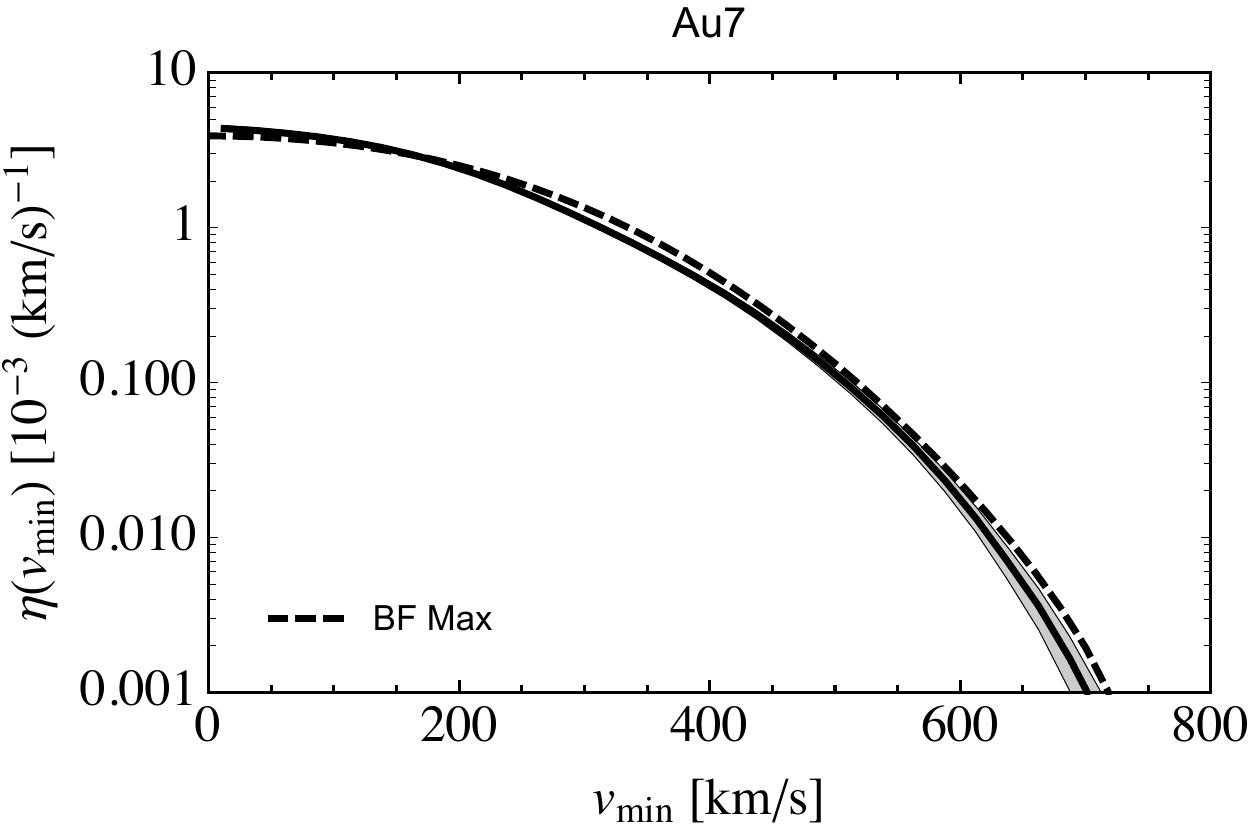}
  \includegraphics[width=0.49\textwidth]{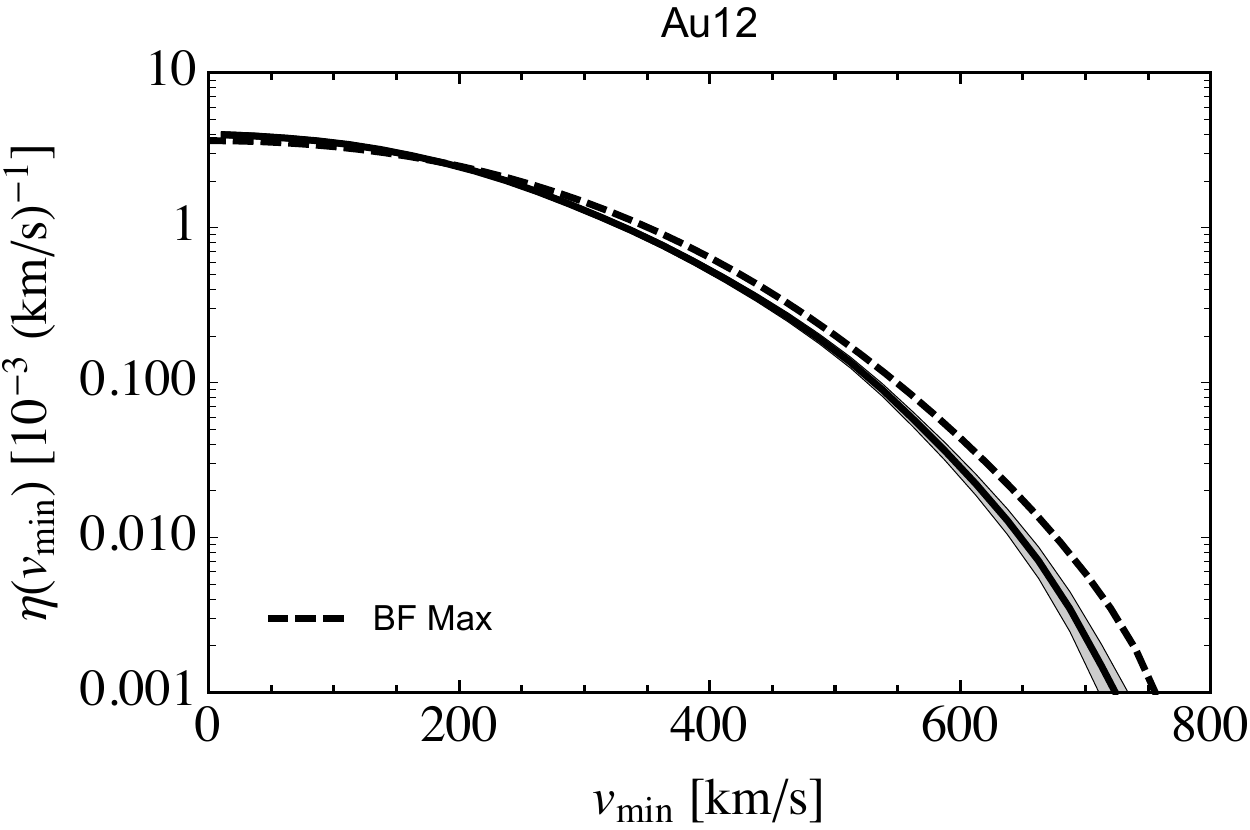}\\
  \includegraphics[width=0.49\textwidth]{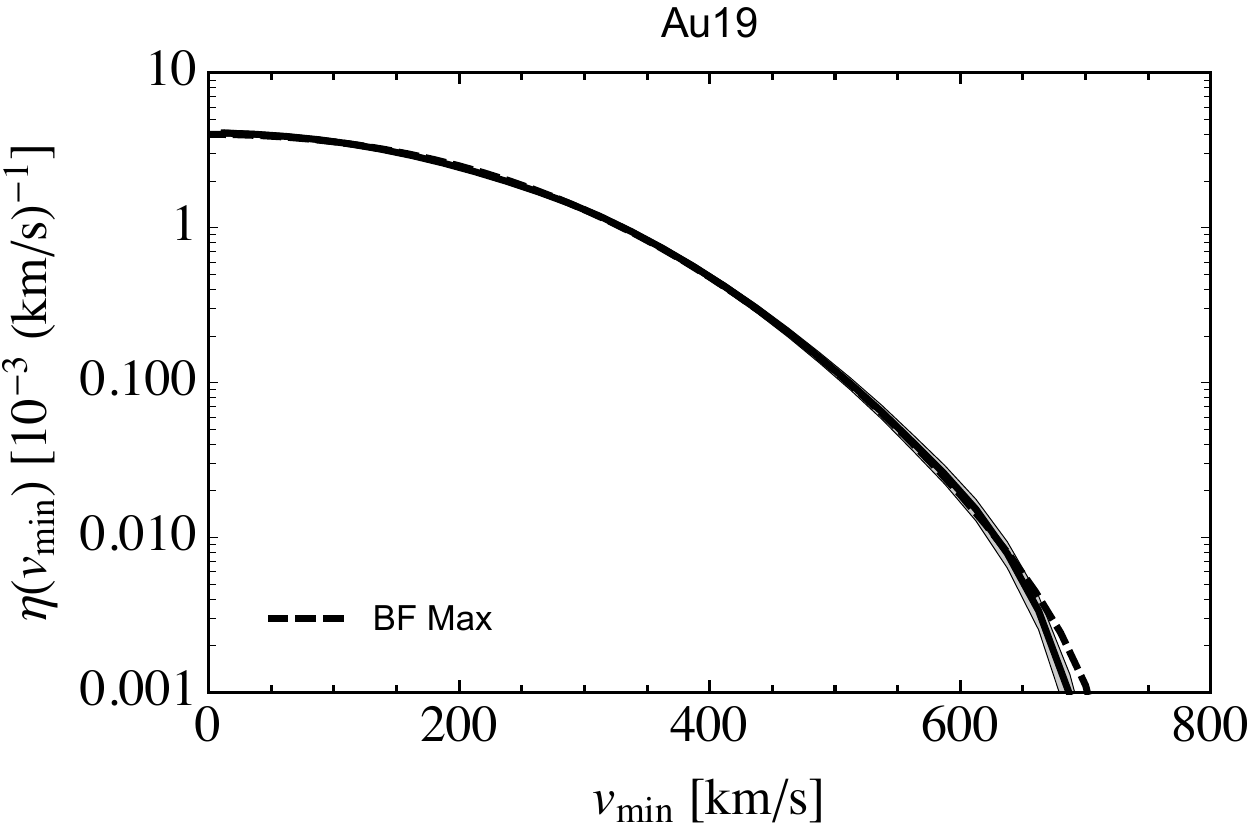}
  \includegraphics[width=0.49\textwidth]{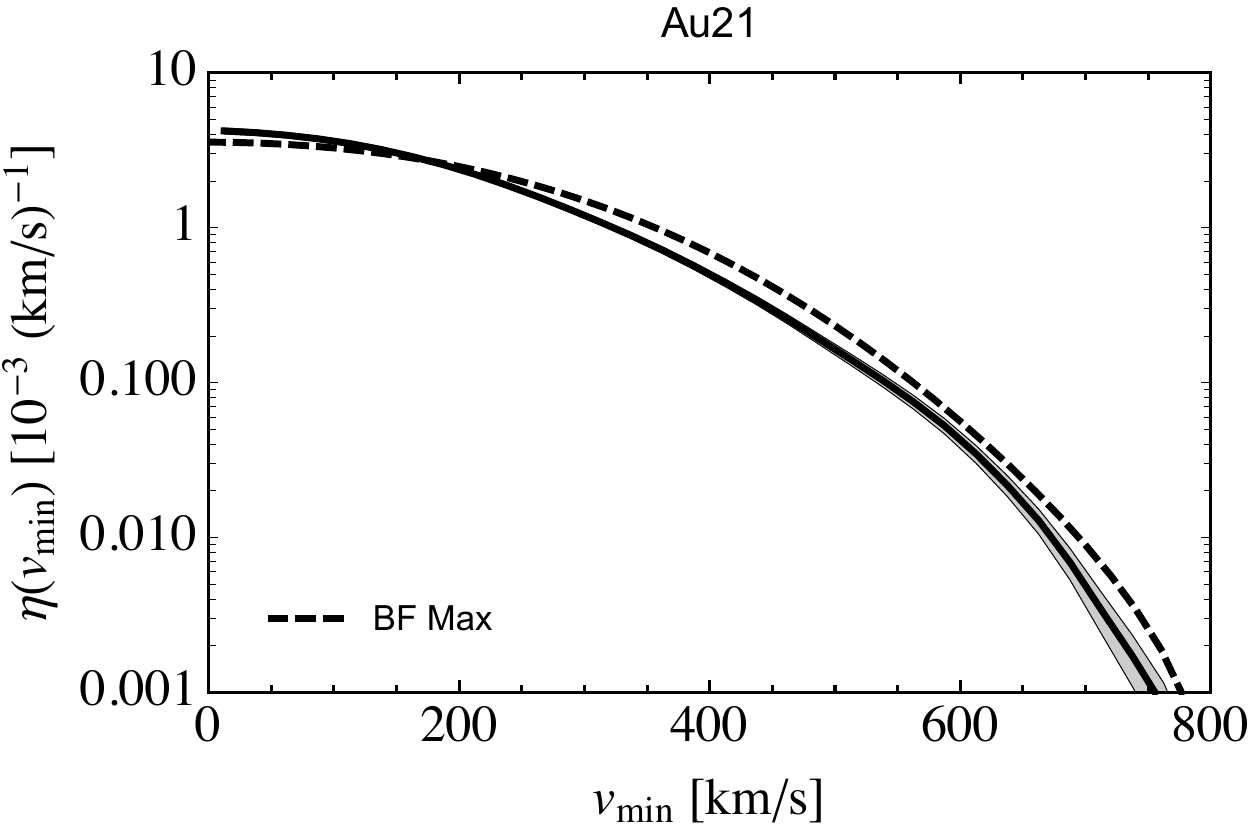}
\caption{Time-averaged halo integrals (solid black lines and shaded bands) of the six Auriga MW-like halos without the GRASP. The solid black lines and the shaded bands correspond to the halo integrals obtained
from the mean DM velocity distribution and the DM velocity distribution at $1\sigma$ from the mean, respectively. The dashed black lines specify the halo integrals obtained from the best fit Maxwellian velocity distribution for each halo.}
\label{fig:eta}
\end{center}
\end{figure}

\begin{figure}[t!]
\begin{center}
\includegraphics[width=0.49\textwidth]{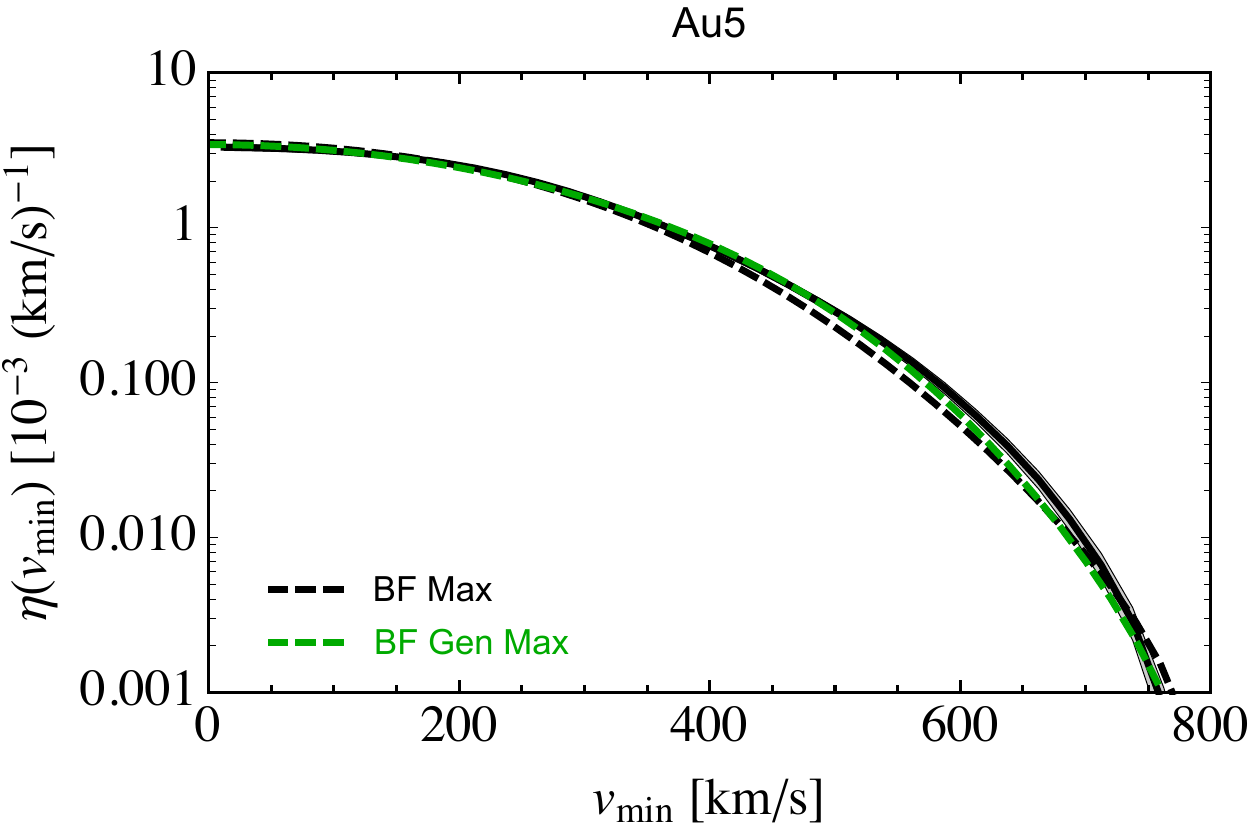}
  \includegraphics[width=0.49\textwidth]{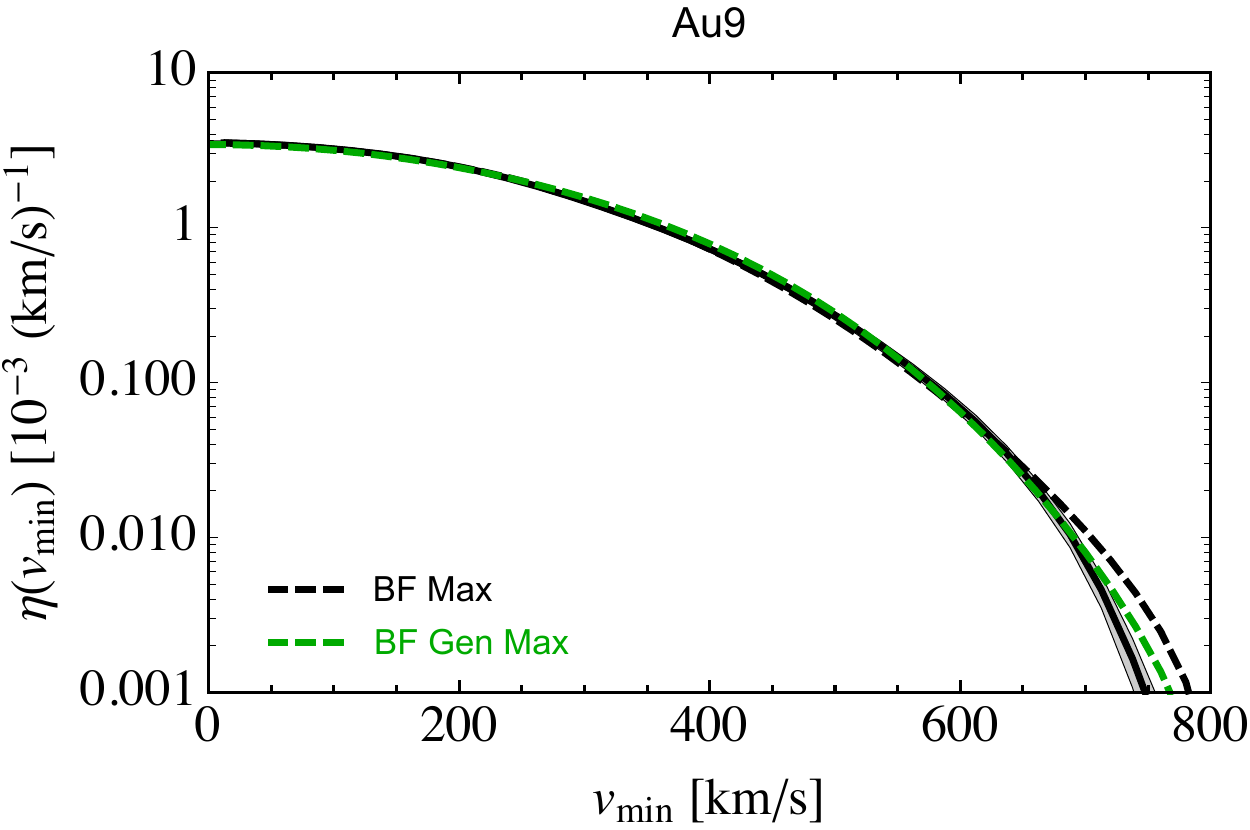}\\
  \includegraphics[width=0.49\textwidth]{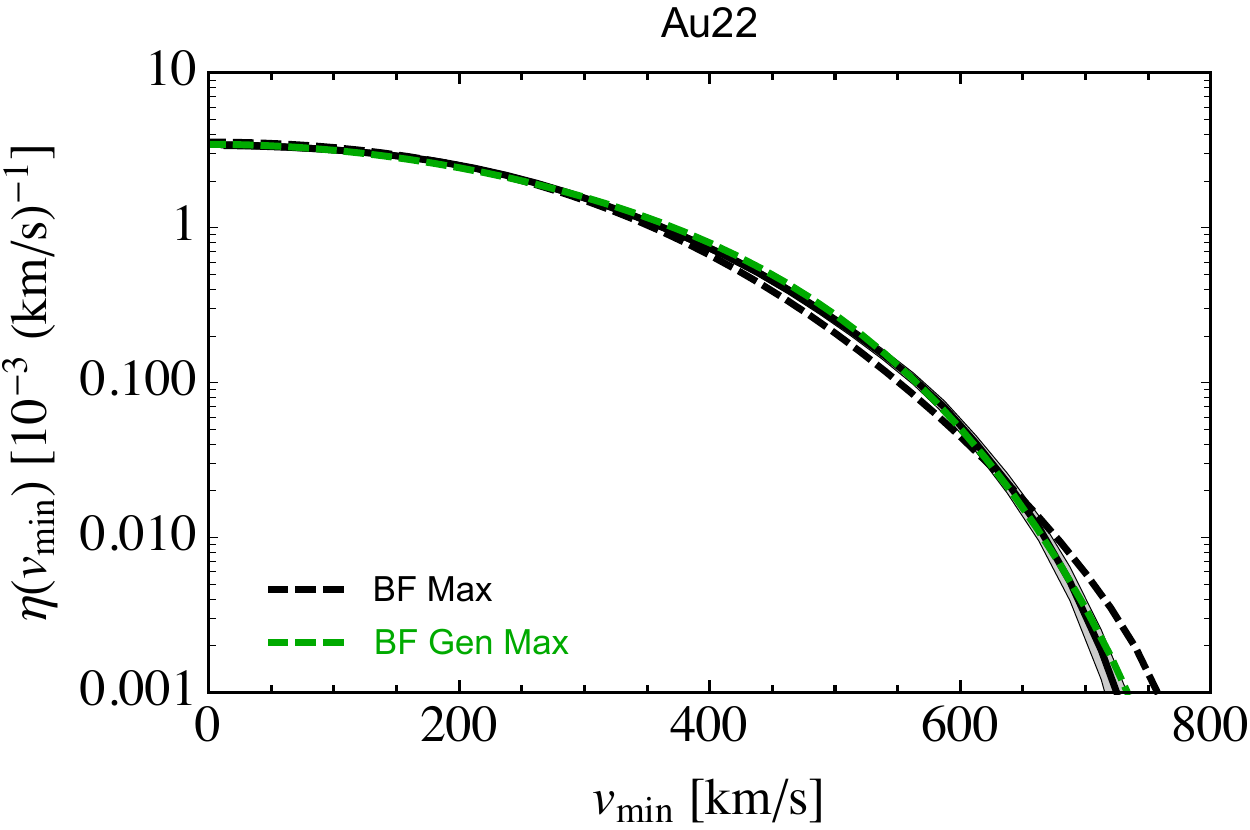}
  \includegraphics[width=0.49\textwidth]{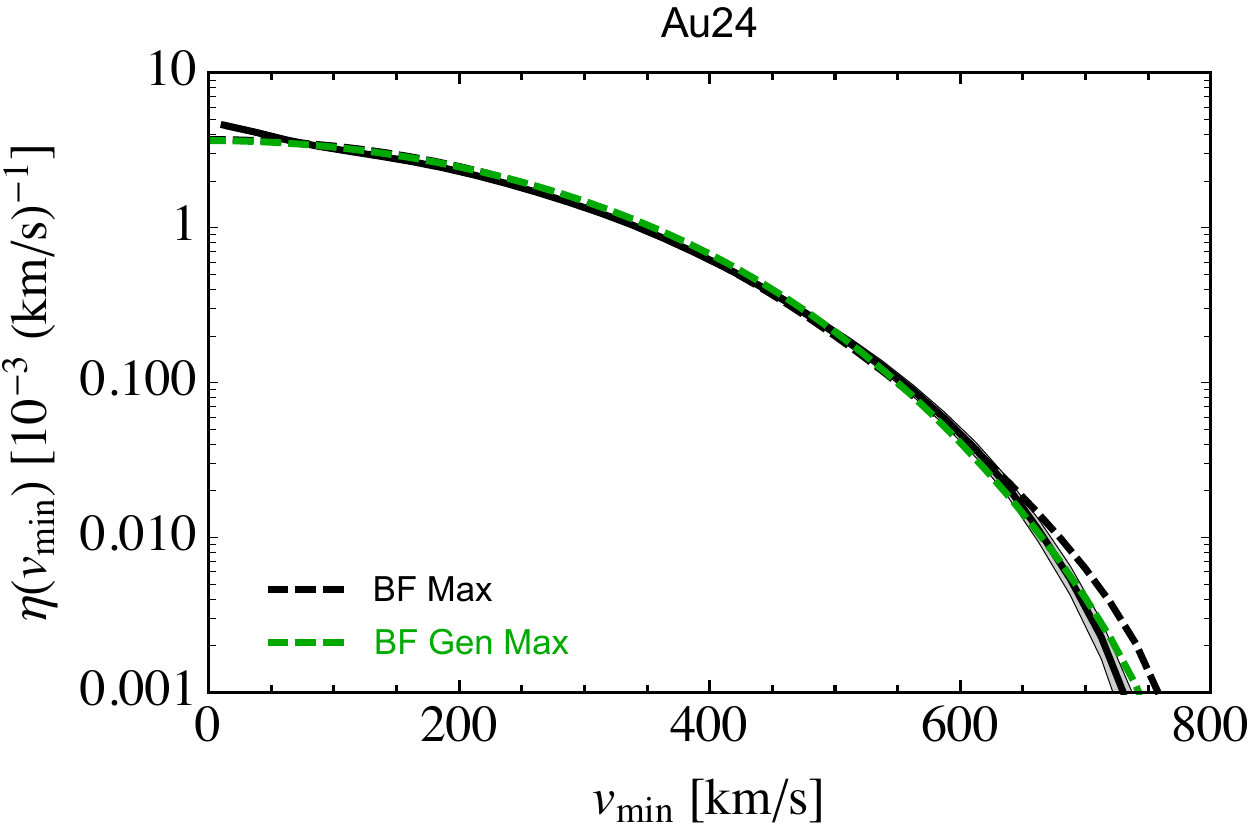}
\caption{Same as figure~\ref{fig:eta-GRASP}, but for the Auriga MW-like halos with the GRASP. The dashed green lines specify the halo integrals obtained from the best fit generalized Maxwellian velocity distribution for each halo.}
\label{fig:eta-GRASP}
\end{center}
\end{figure}

In figures \ref{fig:eta} and \ref{fig:eta-GRASP} we present the time-averaged halo integrals as a function of the minimum speed, $v_{\rm min}$,  obtained from the local DM velocity distributions of the MW-like halos  without and with the GRASP, respectively. The black  solid lines are the halo integrals computed from the mean value
of the local DM velocity distribution, and the shaded bands are obtained by adding and subtracting
one standard deviation to the mean velocity distribution. The dashed black curves specify the halo integrals obtained from the best fit Maxwellian speed distributions for each halo, while the green dashed lines in figure \ref{fig:eta-GRASP} are the halo integrals obtained from the best fit {\it generalized} Maxwellian distribution for the halos with the GRASP.

As seen in figure \ref{fig:eta}, from the six halos without the GRASP, four show a very good agreement between their halo integral and their best fit Maxwellian halo integral. However, the best fit Maxwellian halo integrals for halos Au12 and Au21 do not fully fall within the $1\sigma$ uncertainty band of the halo integrals obtained for those halos. On the other hand, as seen in figure \ref{fig:eta-GRASP} the halo integrals of the four halos with the GRASP show less of an agreement with their best fit Maxwellian halo integrals. In particular, in three out of the four halos with the GRASP, there is a deficit in the high speed tails of the halo integrals compared to their best fit Maxwellian. The best fit generalized Maxwellian halo integrals, however, show very good agreement with the halo integrals of the halos with the GRASP.

Hence, we can conclude that for Auriga MW-like simulated halos with the GRASP, the halo integrals obtained from a generalized Maxwellian velocity distribution (eq.~\eqref{eq:genMax}) with best fit parameters given in table~\ref{tab:BFMax}, agree well with the halo integrals of the simulated halos. 

\begin{figure}[t!]
\begin{center}
 \includegraphics[width=0.48\textwidth]{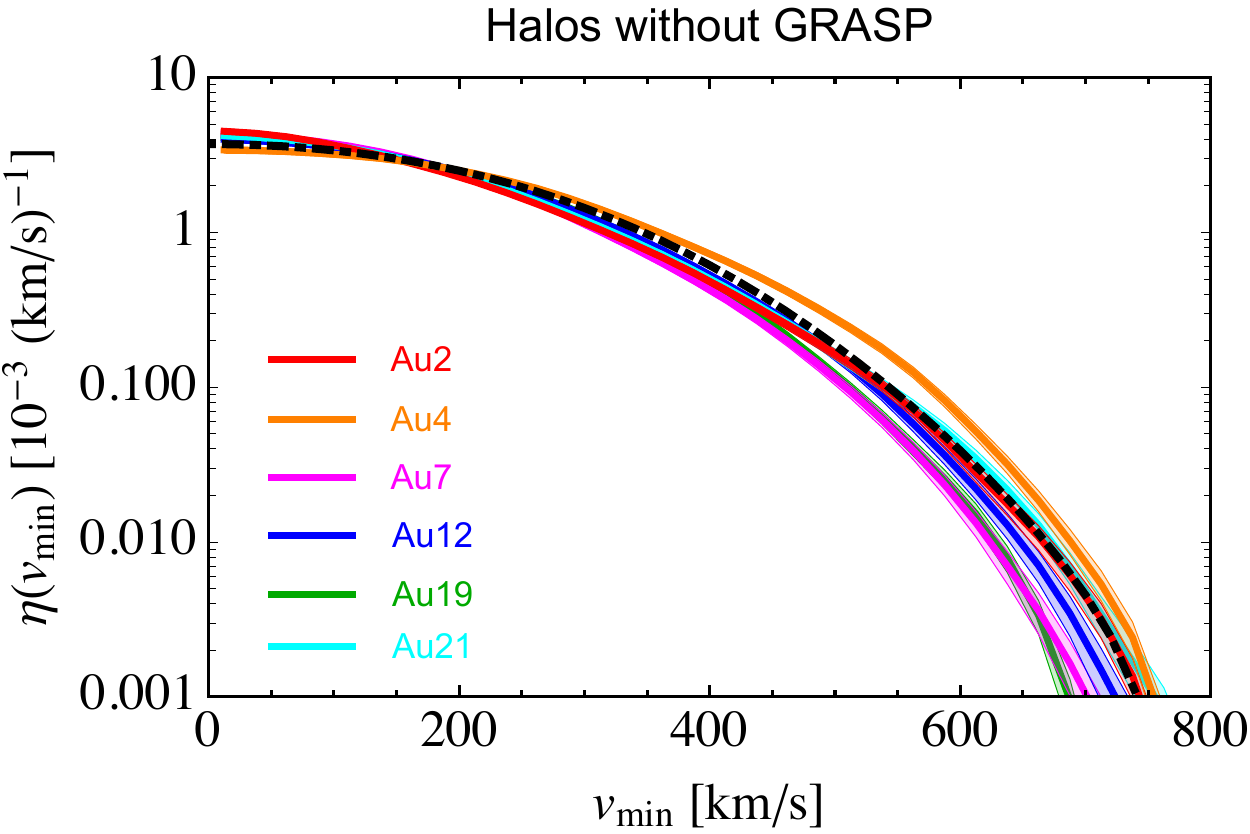}
  \includegraphics[width=0.48\textwidth]{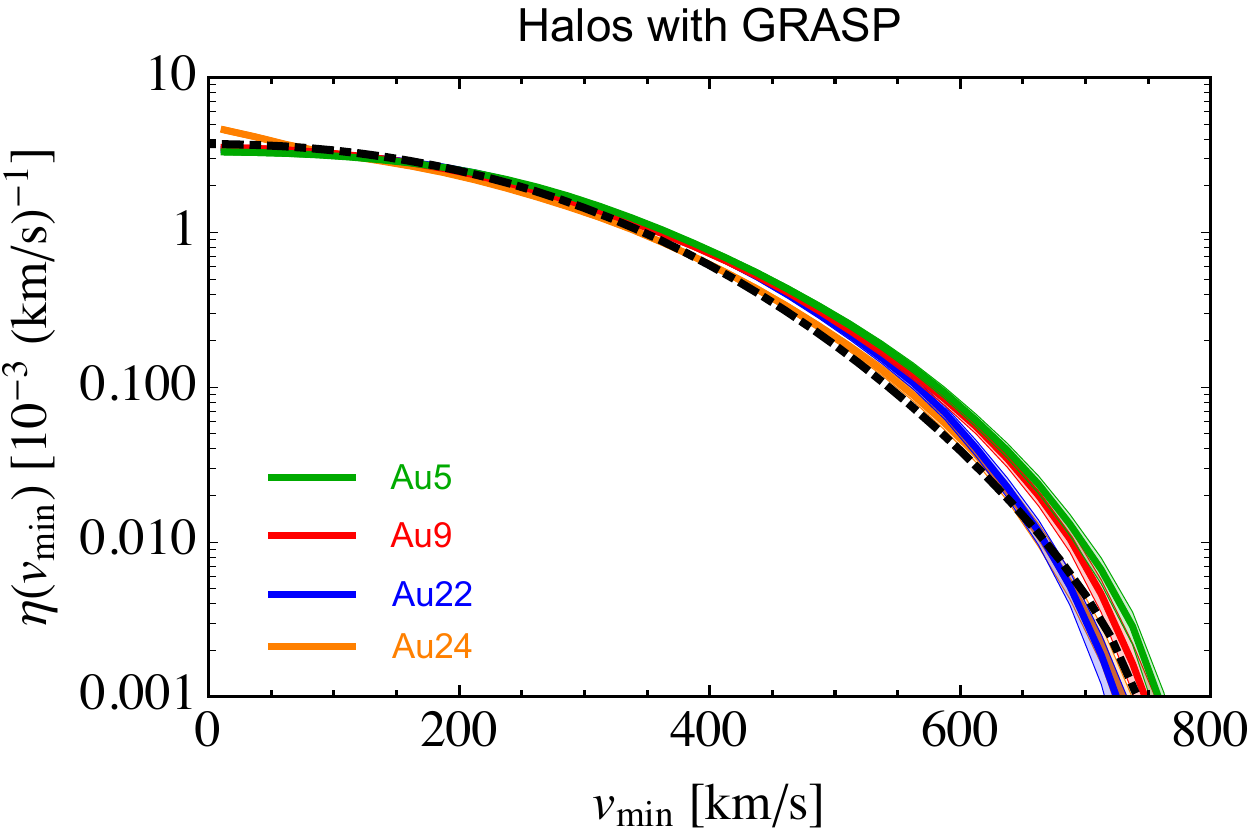}
\caption{A comparison of the time-averaged halo integrals of the Auriga MW-like halos without (left panel) and with (right panel) the GRASP (shaded solid lines and colour bands) and the SHM Maxwellian halo integral (black dot-dashed line). }
\label{fig:etaGRASP-SHM}
\end{center}
\end{figure}

In figure~\ref{fig:etaGRASP-SHM}, we show a comparison of the halo integrals of the MW-like halos without (left panel) and with (right panel) the GRASP (shown also in figures~\ref{fig:eta} and \ref{fig:eta-GRASP}, respectively), and the SHM Maxwellian halo integral obtained from the Maxwellian velocity distribution with a peak speed of 220~km~s$^{-1}$. Clearly, the halo integrals of the simulated halos with and without the GRASP are different from the SHM halo integral, and show some halo-to-halo variation, especially in their high speed tails. The halo integrals for halos with the GRASP  mostly show an excess compared to the SHM halo integral, for $v_{\rm min} \gtrsim 400$~km~s$^{-1}$. Halos Au22 and Au24, however show a deficit compared to the SHM for $v_{\rm min} \gtrsim 650$~km~s$^{-1}$. A larger variation is observed in the tails of the halo integrals of the halos without the GRASP, since they are more numerous. 

%*******************************%

\subsection{Effect on exclusion limits}
\label{sec:exclimits}

As implied by eq.~\eqref{eq:difrate}, variations in the halo integral can have a large impact on the expected number of DM recoils in a direct detection experiment. For a given experimental setup, this results in different interpretations of the DM mass and scattering cross section. To illustrate how the local DM distribution obtained from the simulated halos can alter this interpretation, we  consider the case of spin-independent scattering and assume equal couplings of DM to protons and neutrons. Therefore, the differential cross section is given by,
\begin{equation}
\frac{d \sigma_T}{d E_R}=\frac{m_T A^2 \sigma_{\chi N}^{\rm SI}}{2 \mu_{\chi N}^2 v^2}~F^2(E_R),    
\end{equation}
where $\sigma_{\chi N}^{\rm SI}$ is the spin-independent DM-nucleon scattering cross section at zero-momentum, $A$ is the atomic mass number of the target nucleus, $\mu_{\chi N}$ is the DM-nucleon reduced mass, and $F(E_R)$ is the nuclear form factor that comes from calculating the nuclear matrix elements. Here we use the results presented in ref.~\cite{Fitzpatrick:2012ix} for the spin-independent form factor\footnote{In the more comprehensive non-relativistic effective field theory basis, the spin-independence form factor is known as the form factor for operator $\mathcal{O}_{1}$.}.

To study the implications of the DM component of the GRASP on direct detection event rates, we simulate the signals in two idealised direct detection experiments inspired by projected detectors with different target nuclei, one germanium and the other xenon based. These target materials show promising future sensitivities and provide a large coverage in $m_{\chi}$, when considered together. 
In particular, low-temperature solid-state detectors such as SuperCDMS, which employ germanium targets are extremely sensitive to low-mass DM \cite{Agnese:2017jvy} (much like technologies that use light gases~\cite{Arnaud:2017bjh}, liquid argon~\cite{Agnes:2018ves} and other crystals as targets~\cite{Abdelhameed:2019hmk}). The xenon detector is inspired by the dual phase time projection chambers~\cite{Cui:2017nnn,Aprile:2018dbl,Akerib:2015cja}, a technique also being pioneered for argon~\cite{Aalseth:2017fik} that covers the high DM mass range and lower cross sections. 

For the germanium based detector, we have simulated two crystal designs,  based on the projected HV and iZIP of SuperCDMS SNOLAB~\cite{Agnese:2016cpb}.
The HV design allows experiments to push the analysis threshold very low, enabling greater sensitivity to masses below $\sim5$~GeV. In our setup, we consider an energy range of [$40-300]$~eV, with a flat background of $10\,\mathrm{keV}^{-1}\, \mathrm{kg}^{-1}\, \mathrm{days}^{-1}$, an exposure of $1.6 \times 10^{4}~\mathrm{kg}\,$~days and a flat efficiency of 85\% as in refs.~\cite{Agnese:2016cpb,Kahlhoefer:2017ddj}. To achieve the corresponding 90\% confidence level (CL) with the expected background, we use the standard likelihood ratio method~\cite{Workgroup:2017lvb}. For the iZIP design, one can discriminate electronic and nuclear recoils, substantially decreasing the expected background at higher recoil energies. To avoid as many background events as possible, we take the energy range [$3-30$]~keV. With an exposure of $2.04 \times 10^{4}~\mathrm{kg}$~days, an efficiency of 75\% and 1 expected background event, the sensitivity for DM masses $m_{\chi}>5$~GeV is improved upon the HV detector. These exposures would be achievable by SuperCDMS~\cite{Agnese:2016cpb} after 5 years of operation during 2020-2024.

The xenon experiment is simulated by considering the energy range [$5-50$]~keV and an exposure of $5.6 \times 10^{6}~\mathrm{kg}$~days with a flat efficiency of 50\% as in refs.~\cite{Akerib:2015cja,Mount:2017qzi}. The expected number of background events in this setup is $\sim 6$ and the expected 90\% CL is calculated as in ref.~\cite{DEramo:2016gos}. Notice that we do not include the coherent neutrino background in our calculations, since the experimental parameters chosen are not expected to be sensitive to such interactions. The exposure we consider for the xenon based detector is expected to be achieved by LZ which will operate during 2020-2025~\cite{Akerib:2015cja}.

\begin{figure}[t!]
\begin{center}
  \includegraphics[width=0.49\textwidth]{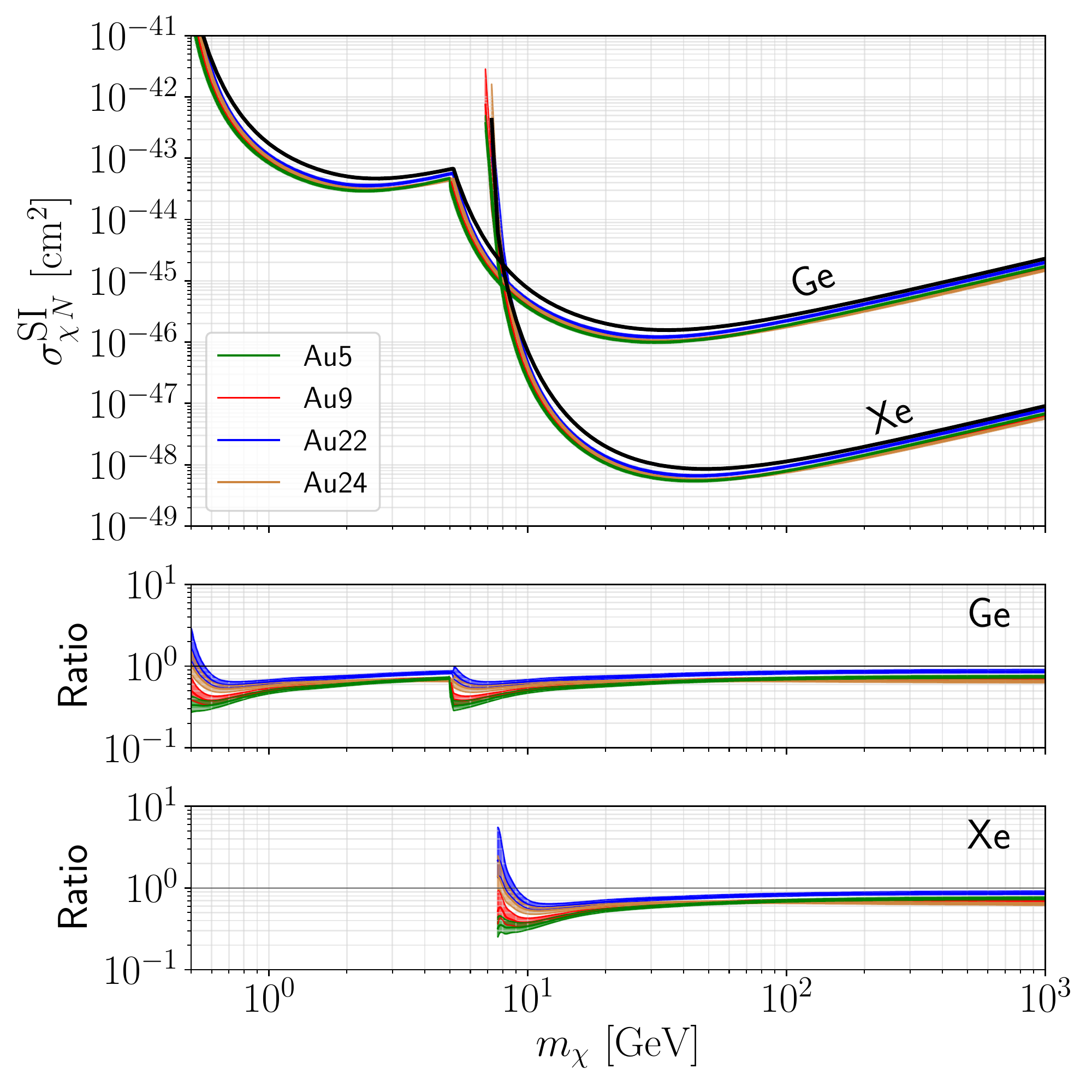}
  \includegraphics[width=0.49\textwidth]{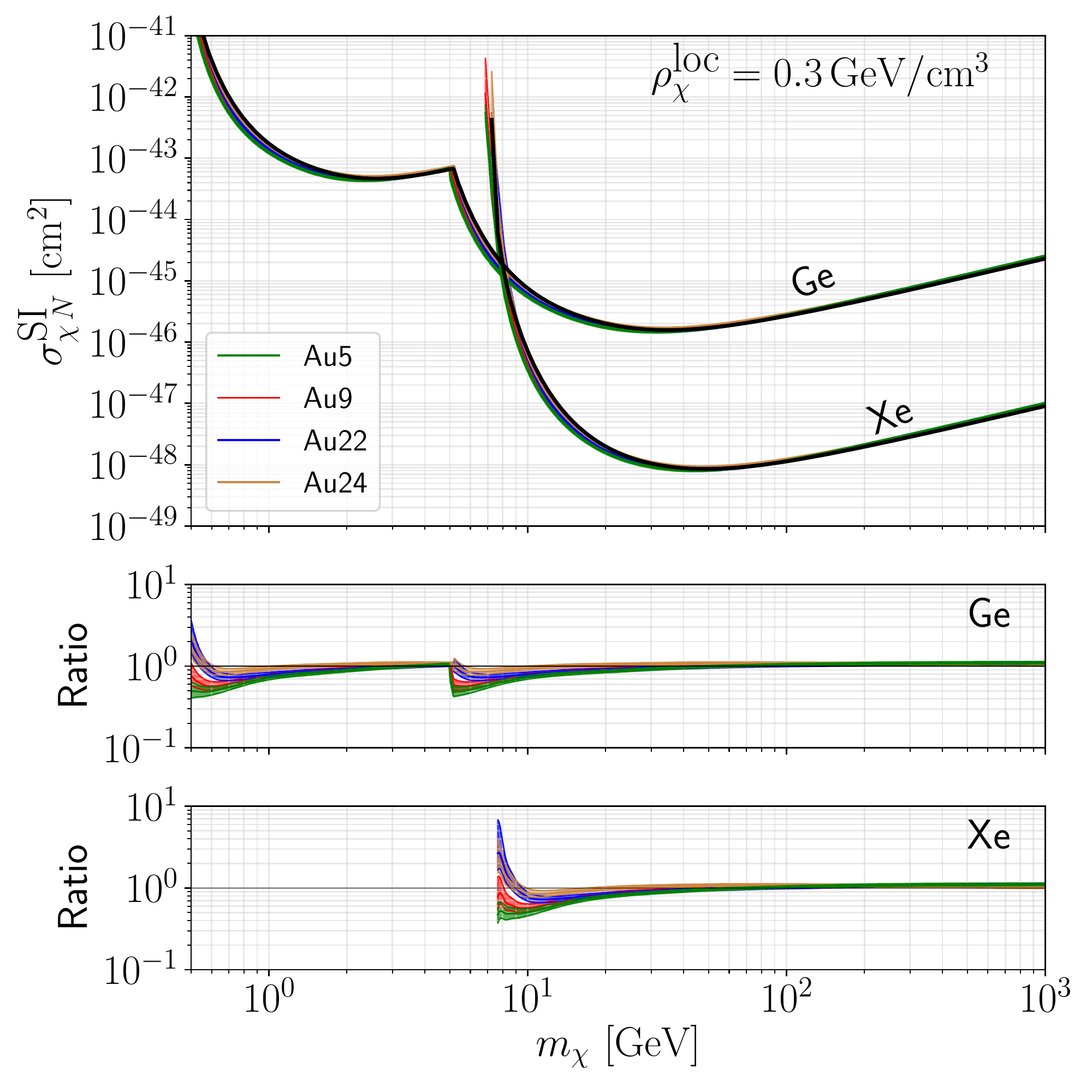}
\caption{Top panels: exclusion limits at 90\% CL from future Ge and Xe experiments in the spin-independent DM-nucleon cross section and DM mass plane for the four MW-like Auriga halos with the GRASP (coloured shaded bands), assuming the local DM density of each halo from table~\ref{tab:Localrho} (left), and fixing $\rho_\chi^{\rm loc}=0.3$~GeV$/$cm$^3$ for all halos (right). The shaded bands in the exclusion limits are obtained from the upper and lower $1\sigma$ limits of the
halo integral for each halo. The black solid curves are the exclusion limits of the SHM Maxwellian. Middle and bottom panels: the ratio of the exclusion limits computed using the local DM distribution of the Auriga simulated halos and the SHM for Ge and Xe, respectively.}
\label{fig:limits_sausage}
\end{center}
\end{figure}

\begin{figure}[t!]
\begin{center}
  \includegraphics[width=0.49\textwidth]{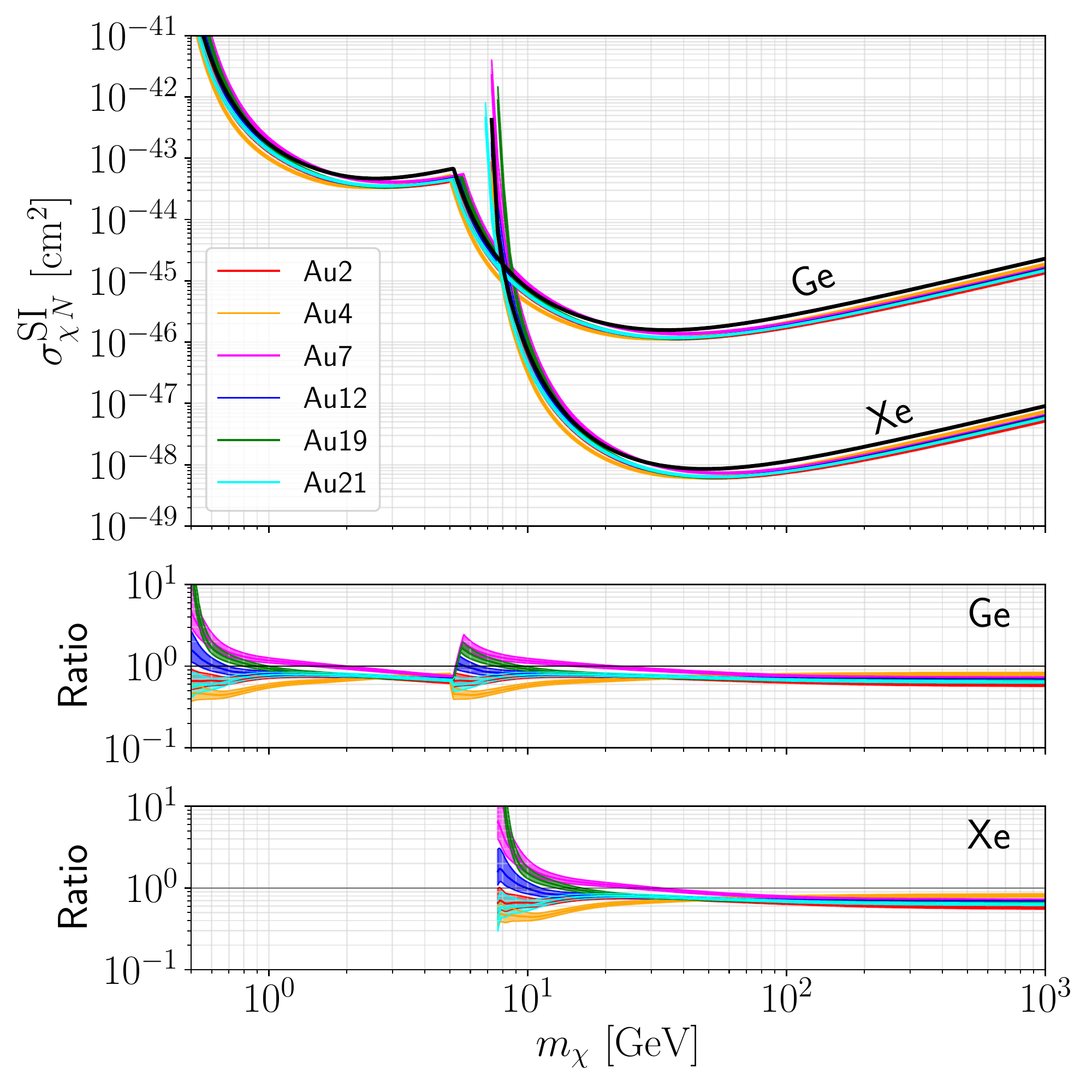}
  \includegraphics[width=0.49\textwidth]{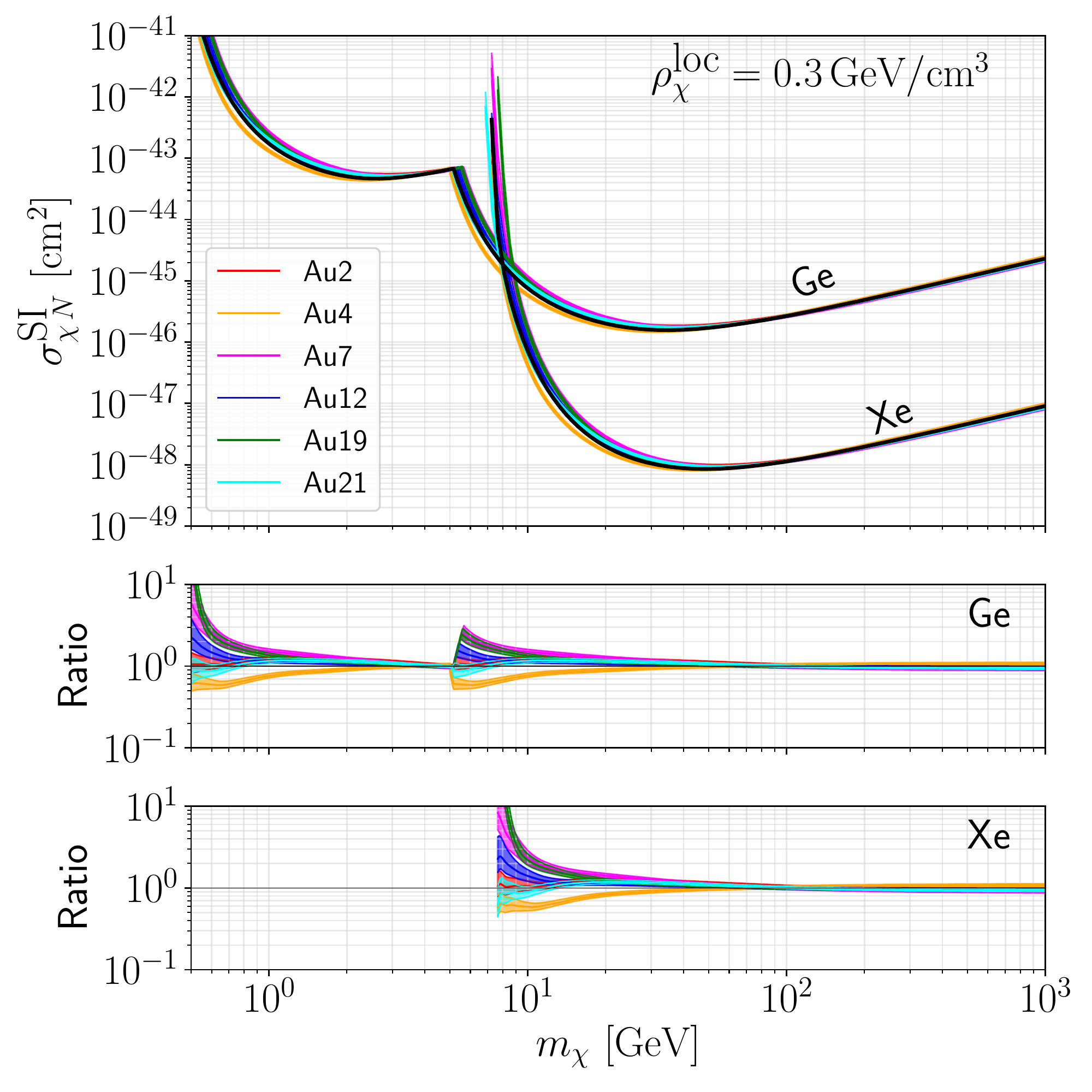}
\caption{Same as figure~\ref{fig:limits_sausage} but for Auriga MW-like halos without the GRASP.}
\label{fig:limits_nonsausage}
\end{center}
\end{figure}

The top panels of figures~\ref{fig:limits_sausage} and \ref{fig:limits_nonsausage} show the exclusion limits at the 90\% CL in the plane of DM mass and spin-independent cross section set by future xenon and germanium experiments using directly the local DM distribution of the Auriga MW-like halos with and without the GRASP, respectively. The exclusion limits are calculated using the halo integrals, $\eta$, shown in figures~\ref{fig:eta} and \ref{fig:eta-GRASP}, where the 1$\sigma$ uncertainty bands on $\eta$ translate to a shaded band around the mean values in the exclusion limits. For comparison, the exclusion limit for the SHM (with a peak speed of 220 km~s$^{-1}$) is also shown as a solid black line. In the left panels of the figure, the local DM density for each halo (from table~\ref{tab:Localrho}) is used in the calculation of the exclusion limits, whereas in the right panels the local DM density is set to 0.3~GeV$/$cm$^3$ for all the simulated halos, to better observe the effect of the velocity distribution\footnote{Strictly speaking, it is inconsistent to change the value of $\rho_{\chi}^{\rm loc}$ for the simulated halos as if it were an independent variable, since this would alter the self-consistency of the halos and would in principle require that $f(v)$ is also varied~\cite{Lavalle:2014rsa}.}.

The two bottom panels of figures~\ref{fig:limits_sausage} and \ref{fig:limits_nonsausage} show the ratio of the exclusion limits computed using the local DM distribution of the Auriga halos and the SHM for each target material, defined by,
\begin{equation}
    \textrm{Ratio}=\frac{\sigma_{\rm 90\%, Auriga}^{\rm SI} (m_\chi)}{\sigma_{\rm 90\%, SHM}^{\rm SI} (m_\chi)},
\end{equation}
where $\sigma_{\rm 90\%, Auriga}^{\rm SI}$ and $\sigma_{\rm 90\%, SHM}^{\rm SI}$ are the limits on $\sigma^{\rm SI}_{\chi N}$ at the 90\% CL at each DM mass for the Auriga halos and SHM, respectively.

Both figures~\ref{fig:limits_sausage} and \ref{fig:limits_nonsausage} show the expected behaviour. In particular, the variation in the local DM density of the simulated halos causes the vertical shift of the exclusion limits with respect to the SHM for all DM masses. The variations in the high speed tail of the halo integrals (shown in figures~\ref{fig:eta} and \ref{fig:eta-GRASP}) lead to variations of the exclusion limits with respect to the SHM at low DM masses. This is because at low DM masses, the experiments probe large $v_{\rm min}$, and therefore only the high speed tail of the halo integrals affects the recoil rate. 

Furthermore, by comparing figures~\ref{fig:limits_sausage} and \ref{fig:limits_nonsausage}, we can see a greater variation in the exclusion limits of the halos without the GRASP than the halos with the GRASP, even when the local DM density is fixed to 0.3 GeV$/$cm$^3$ for all halos. This is because the halos without the GRASP are more numerous, with a greater variation in the peak speeds of their DM velocity distributions, resulting in a larger variation in the tails of their halo integrals. This results in a more pronounced halo-to-halo variation in their exclusion limits at low DM masses. 

As it can be seen from the right panel of figure~\ref{fig:limits_sausage}, for DM masses below $\mathcal{O}$(10 GeV) the exclusion limits for the simulated halos with the GRASP show a mild shift towards smaller masses compared to the SHM. This is due to the excess observed in the tails of the halo integrals for halos with the GRASP compared to the SHM halo integral (as seen in right panel of figure~\ref{fig:etaGRASP-SHM}). While the right panel of figure~\ref{fig:limits_nonsausage} shows that the exclusion limits for the simulated halos without the GRASP are either shifted towards smaller or larger DM masses with respect to the SHM, depending on the halo.

Notice that the results will be the same for any  velocity-independent DM-nucleus interaction. For such interactions, the event rate is proportional to the halo integral given in eq.~\ref{eq:halointegral}, and any variations in the halo integral leads to the same variations in the direct detection exclusion limits. The spin-dependent interaction is a common example of such an interaction. 

Finally, we compare our results with those of ref.~\cite{Buch:2019aiw}, which finds that for DM masses below $\sim 10$~GeV the DM component of the GRASP leads to significantly weaker direct detection limits compared to the SHM. This is due to the specific DM speed distribution adopted by ref.~\cite{Buch:2019aiw} (based on the results of refs.~\cite{Necib:2018iwb, Necib:2018igl}) with a peak speed shifted to lower speeds compared to the SHM, and consequently halo integrals with tails shifted to smaller $v_{\rm min}$. The local DM distribution of the simulated Auriga halos with the GRASP, however, show the opposite behaviour, leading to mildly stronger direct detection constraints at low DM masses, compared to the SHM.

%*******************************%

\section{Conclusions}
\label{sec:conclusions}

In this work we have studied the DM component of the Gaia Radially Anisotropic Stellar Population (GRASP) in a set of magneto-hydrodynamical simulations of MW-like galaxies of the Auriga project~\cite{Grand:2016mgo}, and investigated its implications for DM direct detection. We first identified MW-like simulated halos by requiring that the total stellar mass of the simulated galaxy should lie in the $3\sigma$ range of the observed MW stellar mass~\cite{McMillan:2011wd}, and the rotation curve of the simulated galaxy should agree with the observed MW rotation curve~\cite{Iocco:2015xga}. We found that 10 halos satisfy our criteria and four of them have the GRASP component. We have extracted the fraction and anisotropy of the DM component of the GRASP in the Solar neighbourhood, as well as the local DM density and velocity distribution of halos with and without the GRASP. Finally, we have simulated the signals in two future germanium and xenon direct detection experiments, and showed how  direct detection exclusion limits are altered for halos with and without the GRASP. We summarize our findings below.

\begin{itemize}

\item The fraction of the DM particles belonging to the GRASP in a torus around the Solar circle is between 0.6\% and 17\% for the Auriga MW-like halos with the GRASP. The anisotropy parameter of these DM particles is in the range of $\beta=[0.48 - 0.82]$. There exists  an anti-correlation between the  fraction and anisotropy of the DM particles belonging to the GRASP in the torus.

\item The local DM density of the MW-like halos with the GRASP is in the range of $\rho_\chi^{\rm loc}=[0.37 - 0.48]$~GeV$/$cm$^3$, depending on the halo. The slope of the logarithmic DM profile in the Solar neighbourhood for halos with a GRASP ranges from $1.66\pm0.022$ to $1.95\pm0.022$.

\item The local DM speed distributions of the MW-like halos with the GRASP are shifted to higher speeds compared to the Maxwellian speed distribution of the Standard Halo Model (SHM).

\item The halo integral obtained from a \emph{generalized Maxwellian} velocity distribution (given in eq.~\eqref{eq:genMax}) falls within the $1\sigma$ uncertainty band of the halo integrals of the simulated MW-like halos with the GRASP. The best fit parameters of the generalized Maxwellian distribution are given in table~\ref{tab:BFMax}. For the four MW-like halos with the GRASP, these parameters are $(v_0~[{\rm km~s}^{-1}],\alpha)=(296, 1.42)$, $(292, 1.37)$, $(319,1.77)$, and $(280, 1.32)$.

\item Variations in the local DM density of the simulated halos with and without the GRASP lead to a vertical shift in the expected exclusion limits in the DM mass {\em vs} cross section plane with respect to the SHM in forthcoming Ge and Xe direct detection. Variations in the high speed tail of the halo integrals of the halos with and without the GRASP are responsible for a horizontal shift in the direct detection exclusion limits at low DM  masses. In particular for DM particle masses below $\mathcal{O}$(10 GeV), the exclusion limits for the halos with the GRASP are slightly shifted towards smaller masses compared to the SHM.

\item There is a larger halo-to-halo variation between the direct detection limits for halos without the GRASP, due to them being more numerous, compared to halos with the GRASP.

\item In general, the effect of the GRASP on direct detection exclusion limits is small, and generally smaller than the intrinsic halo-to-halo variation.
    
\end{itemize}

Finally, we compare our findings with those in refs.~\cite{Necib:2018iwb} and \cite{Evans:2018bqy}. The general behaviour of the local DM speed distributions for the halos with the GRASP compared to the SHM is shown in figure~\ref{fig:fvGRASP-SHM}, where the DM speed distributions are shifted to higher speeds compared to the SHM. This result is at odds with the findings of ref.~\cite{Necib:2018iwb}, where  the DM distribution was found to be shifted to lower speeds compared to the SHM. Ref.~\cite{Evans:2018bqy} suggests  that a linear combination of an isotropic and an anisotropic Gaussian distribution can be used for the local DM velocity distribution. We find however, that such a combination does not fit the local DM velocity distributions of the simulated halos with the GRASP, and instead a  generalized Maxwellian distribution provides a better fit. More specifically, a generalized Maxwellian distribution fits well the DM halo integrals of the simulated MW-like halos with the GRASP.

Notice that, in this work, we have not considered the impact of  structures which are below the resolution limit of our cosmological simulations, or those which require simulating their exact location and motion in the halo. Examples of the latter are the Sagittarius dwarf galaxy and the Large Magellanic Cloud, which could perturb the local DM density or velocity distribution~\cite{2019ApJ...879L..15H, Besla:2019xbx}, and whose detailed effects can only be studied with specially designed simulations. The GRASP, the structure we have studied in this work, is one possible source of astrophysical uncertainty in the interpretation of DM direct detection results. Structures such as the Sagittarius dwarf and the Large Magellanic Cloud introduce additional astrophysical uncertainties. Future astronomical data together with high resolution simulations specifically modeling the orbits of these objects will lead to a better understanding of such uncertainties.

%*******************************%
\subsection*{Acknowledgements}
We thank Wyn Evans and Annika Peter for helpful comments on the manuscript, and Christopher McCabe for discussions. NB and AF acknowledge support by a European Union COFUND/Durham Junior Research Fellowship (under EU grant agreement no.~609412). NB thanks the Erwin Schr\"odinger International Institute for hospitality while part of this work was completed. NB is grateful to the Institute for Research in Fundamental Sciences in Tehran for their hospitality during her visit. NB has received support from the European Union's Horizon 2020 research and innovation programme under the Marie Sklodowska-Curie grant agreement No.~690575. This work was supported by the Science and Technology Facilities Council 
(STFC) consolidated grant ST/P000541/1. CSF acknowledges support by the 
European Research Council (ERC) through Advanced Investigator grant DMIDAS 
(GA 786910). FAG acknowledges financial support from CONICYT through the project FONDECYT Regular Nr. 1181264, and funding from the Max Planck Society through a Partner Group grant. FM is supported by the Program ``Rita Levi Montalcini" of the Italian MIUR. This work used the DiRAC Data Centric system at Durham 
University, operated by the Institute for Computational Cosmology on 
behalf of the STFC DiRAC HPC Facility (\url{www.dirac.ac.uk}). This 
equipment was funded by BIS National E-infrastructure capital grant 
ST/K00042X/1, STFC capital grant ST/H008519/1, and STFC DiRAC Operations 
grant ST/K003267/1 and Durham University. DiRAC is part of the National 
E-Infrastructure.

\bibliographystyle{JHEP}
\bibliography{./refs}

\end{document}